\documentclass[twocolumn,final,superscriptaddress,longbibliography,aps,10pt]{revtex4-2}

\usepackage[utf8]{inputenc}
\usepackage{calc}
\usepackage{graphicx}
\usepackage{amsmath,amssymb,amsthm}
\usepackage{mathtools}
\usepackage{txfonts}
\usepackage{bm}
\usepackage{color}
\usepackage{hyperref}
\usepackage{multirow}

\newcommand{\avg}[1]{\left\langle #1 \right\rangle}

\begin{document}

\author{Giulio Burgio}
\email{giulioburgio@gmail.com}
\affiliation{Departament d'Enginyeria Inform\`atica i Matem\`atiques, Universitat Rovira i Virgili, 43007 Tarragona, Spain}
\affiliation{Vermont Complex Systems Institute, University of Vermont, Burlington, VT 05405}

\author{Guillaume St-Onge}
\affiliation{Department of Physics, the Roux Institute at Northeastern University, Portland, ME 04101}
\affiliation{Network Science Institute, the Roux Institute at Northeastern University, Portland, ME 04101}
\author{Laurent H\'{e}bert-Dufresne}
\affiliation{Vermont Complex Systems Institute, University of Vermont, Burlington, VT 05405}
\affiliation{Department of Computer Science, University of Vermont, Burlington, VT 05405}

\title{Characteristic scales and adaptation in higher-order contagions}

\begin{abstract}
People organize in groups and contagions spread across them. A simple stochastic process, yet complex to model due to dynamical correlations within and between groups. Moreover, groups can evolve if agents join or leave in response to contagions. To address the lack of analytical models that account for dynamical correlations and adaptation in groups, we introduce the method of generalized approximate master equations. We first analyze how nonlinear contagions differ when driven by group-level or individual-level dynamics. We then study the characteristic levels of group activity that best describe the stochastic process and that optimize agents' ability to adapt to it. Naturally lending itself to study adaptive hypergraphs, our method reveals how group structure unlocks new dynamical regimes and enables distinct suitable adaptation strategies. Our approach offers a highly accurate model of binary-state dynamics on hypergraphs, advances our understanding of contagion processes, and opens the study of adaptive group-structured systems. \\

\textbf{Note:} This is the author-accepted manuscript version of a work published in \textit{Nature Communications}. The final published version is available at: \url{https://doi.org/10.1038/s41467-025-59777-0}.

\end{abstract}

\maketitle

\section*{Introduction}

Recent studies on higher-order contagions have largely consisted of model development efforts. These models focus on nonlinear transmission mechanisms defined at the level of single groups, meaning the groups a susceptible individual belongs to influence the latter independently from each other~\cite{ferraz2024contagion,wang2024epidemic}.
The nonlinear effects are important given the experimental evidence that some social contagions can benefit from network clustering or group structure, an effect expected in superlinear and threshold-like contagions (so-called `complex contagions')~\cite{centola2010spread}. However, other empirical results have shown that the distribution and correlations of contacts across an individual's multiple groups can shape contagions as much as the amount of infected contacts within any given group~\cite{rogers2003diffusion, watts2007influentials}. In fact, having---say---ten infected contacts evenly scattered over ten groups may have more impact than having all ten infected contacts in a single group~\cite{ugander2012structural}. There is thus a need for models of higher-order contagions able to capture heterogeneity and correlations, both within and across groups; yet, these are ignored by the current modeling trends.

There are multiple network effects either at the node and group levels that can affect a contagion. Network neighbors are very different from random members of a population and often from each other. Any heterogeneity implies that sampling a random connection is different from sampling a random node, which is the statistical bias behind the friendship paradox where ``your friends have more friends than you do.'' Accurate descriptions of dynamics on networks therefore rely on capturing important heterogeneities in how dynamical processes see a networked population. Because of their connectivity or degree, not all nodes in the network follow the same dynamics, and neither do all nodes of the same degree because of their different neighborhoods. The same is true in higher-order networks---``your friends belong to more groups than you do’’---, but the states of group members are also more correlated than expected at random. Ignoring these effects can lead to erroneous conclusions about how networks support dynamics, since degree heterogeneity~\cite{st-onge2022influential} and dynamical correlations~\cite{burgio2021network,burgio2024triadic} shape both critical and noncritical behaviors.

In this paper, we capture group effects analytically and explore three questions around the characteristic scales of contagions on higher-order networks.
First, how different are nonlinear contagions operating at the level of groups---based on how many members of a group are infectious---from those operating at the level of nodes---based on how many neighbors are infectious. Because the latter is technically more involved than recent models of hypergraph contagions, we introduce a new mathematical tool called \textit{generalized approximate master equations}, which, by accounting for the activity around a node, captures dynamical correlations within and---crucially---across groups, allowing us to address our questions precisely.
The amount of possible activity configurations, however, grows exponentially with both the number and the size of the groups incident on a node. We circumvent this problem by introducing an activity scale that keeps the representation scalable, while leading to a second, more technical question: What is the scale of group activity a modeler should consider to best predict the true, stochastic dynamics?

Lastly, we ask a similar question from a more applied perspective: What is the activity scale agents should rely on to adapt and, for example, avoid a contagion? In fact, capturing intra- and inter-group correlations brings a huge bonus with it: it allows for the study of adaptive behavior on higher-order networks, i.e., \textit{adaptive hypergraphs}; a possibility---to the best of our knowledge---excluded until now. Relevant adaptive responses as those observed in social, cultural, and economic systems~\cite{gross2009adaptive}, indeed rely on the fact that agents have some valuable knowledge of the state of the groups around them (and beyond); otherwise, adaptation is reduced to a blind rewiring of connections.
A further question thus finally arises: Is there any relationship between the most appropriate activity scale for modelers trying to predict a contagion and for agents trying to adapt to it?

\section*{Results}

\subsection*{Generalized approximate master equations}
\label{sec:game}

To accurately capture dynamical correlations and the local state of the dynamics, mathematical models often rely on approximate master equations (AME)~\cite{house2008deterministic, hebert2010propagation,marceau2010adaptive,lindquist2011effective,gleeson2013binary, o2015mathematical,unicomb2019reentrant, unicomb2021dynamics, kim2023contagion, st-onge2021master}. One can describe contagion dynamics on networks with communities or higher-order structures by distinguishing nodes by their state (infectious or susceptible) and membership (how many groups they belong to) as well as groups by their size (how many nodes they contain) and composition (how many infectious nodes they contain)~\cite{hebert2010propagation}. Or, one can describe contagion dynamics on random (pairwise) networks by distinguishing nodes by their number of neighbors and infectious neighbors~\cite{marceau2010adaptive}. These are sometimes called group- and node-based AME, respectively. The former take into account the effects of groups on contagion dynamics but fall back on heterogeneous pairwise approximation~\cite{eames2002modeling} when describing random networks. The latter retain dynamical correlations with high accuracy on random networks, but cannot account for group structure.

\begin{figure}
    \centering
    \includegraphics[width=\linewidth]{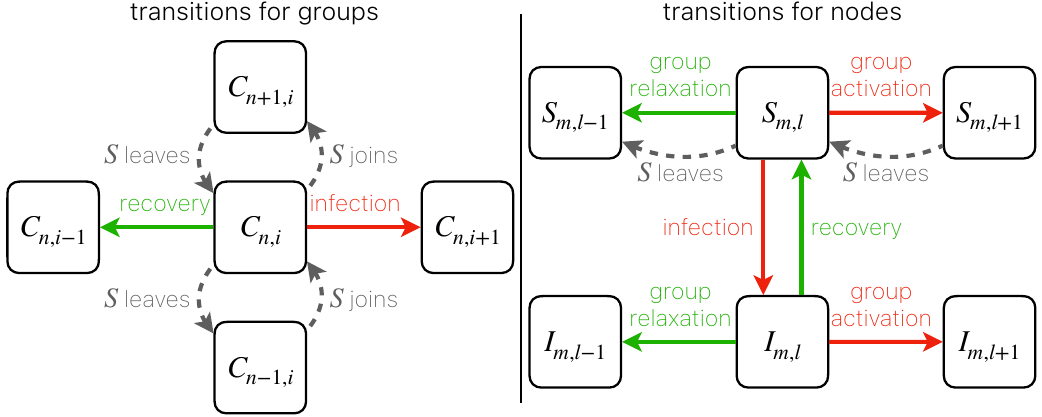}
    \caption{States and transitions for generalized approximate master equations. Groups are identified by their size $n$ and number of infectious members $i$. Nodes are identified by their group membership $m$ and the number $l$ of those that are considered active (i.e., containing at least $\bar{i}$ infected members). GAME transitions, Eqs.~(\ref{eq:master_equations}), correspond to binary-state ($\{S,I\}$) contagion dynamics and indicated by red and green arrows. Adaptive GAME transitions, Eqs.~(\ref{eq:rewiring}), are indicated by dashed arrows.}
    \label{fig:compartments}
\end{figure}

Compared to standard heterogeneous mean-field models~\cite{pastor-satorras2015epidemic,landry2020effect}, AME stand out by tracking the full distribution of dynamical states within mesoscopic units---i.e., the number of infectious nodes within either a group or the neighborhood of a node---instead of just average states.
Recent results have shown that these state distributions can be very heterogeneous and even bimodal, explaining blind spots of standard mean-field approaches that fail to predict important dynamical regimes~\cite{st-onge2021social,st-onge2021master}. 

In order to tackle our research questions, we aim to suitably combine node- and group-based AME approaches by replacing the weakness of one with the strength of the other.  Let us consider a generic binary-state dynamics on infinite-size random hypergraphs, i.e., networks with groups.
Nodes can either be susceptible ($S$) or infected ($I$) and have a membership $m$, defined as the number of groups incident to a node, drawn from the probability distribution $\{g_m\}_{m\in\mathbb{N}}$. Groups are of various size $n$ drawn from the probability distribution $\{p_n\}_{n\in\mathbb{N}}$. We partition groups and nodes according to their local properties. Specifically, we track $C_{n,i}(t) \in [0,p_n]$, the proportion of groups of size $n$ with $i \in \lbrace 0, \dots, n \rbrace$ infected nodes at time $t$.
We also track $S_{m,l}$ and $I_{m,l} \in [0,g_m]$, the fraction of susceptible and infected nodes with membership $m$ and $l \in \lbrace 0, \dots, m \rbrace$ incident active groups. For a given node, any group to which it belongs is labeled \textit{active} if it contains at least $\bar{i}$ infected nodes other than the focal node.
Accordingly, we call $l$ the active membership of a node. The dynamical variables above are also to be interpreted as joint probabilities, i.e., $C_{n,i} \equiv \textrm{Prob}(n,i)$, $S_{m,l} \equiv \textrm{Prob}(S,m,l)$, and $I_{m,l} \equiv \textrm{Prob}(I,m,l)$, with normalizations $\sum_{n,i} C_{n,i} = \sum_n p_n = 1$ and $\sum_{m,l} (S_{m,l} + I_{m,l}) = \sum_m g_m = 1$.
Unless specified, sums run over all possible values. Note that, in this framework, group-based AME are recovered as the degenerate case with $\bar{i} \geqslant n_{\textrm{max}}$---the maximal group size---, while node-based AME are obtained by setting $p_n = \delta_{n,2}$ and $\bar{i} = 1$.

Introducing the parameter $\bar{i}$ is a convenient way to detect where activity is more localized in the system, i.e., around which nodes and groups based on their membership and size, respectively. In particular, the states of adjacent groups are correlated through (at least) the states of the members they share, thus breaking the assumption---made in heterogeneous mean-field and group-based AME models---that the activity around a node scales linearly with its membership. By splitting groups based on $\bar{i}$ we no longer neglect those inter-group correlations, while still keeping the computational cost of the model low. In this regard, observe that a straight merge of group- and node-based AME would require to track the number of incident groups with any given number of active nodes (i.e., $l_1$ groups with one active node, $l_2$ groups with two active nodes, etc.), leading to a combinatorial explosion of node classes, hence equations. As we will show, the coarse-grained partition based on $\bar{i}$ turns out to be a parsimonious approach.

Let us start by defining the \textit{effective} infection (recovery) rate $\bar{\beta}_{n,i}$ ($\bar{\alpha}_{n,i}$) for a node within a group of size $n$ with $i$ infected members and the effective infection (recovery) rate $\tilde{\beta}_{m,l}$ ($\tilde{\alpha}_{m,l}$) for a node of membership $m$, $l$ of which are active.
From these definitions, we introduce the following set of \textit{generalized approximate master equations} (GAME), schematized in Fig.~\ref{fig:compartments},
\begin{subequations}
\label{eq:master_equations}
\begin{align}
    \frac{\mathrm{d}C_{n,i}}{\mathrm{d}t} =& \;\bar{\alpha}_{n,i+1} (i+1) C_{n,i+1} - \bar{\alpha}_{n,i} i C_{n,i} \notag \\
                                           & + \bar{\beta}_{n,i-1} (n-i+1) C_{n,i-1} - \bar{\beta}_{n,i} (n-i) C_{n,i} \label{eq:master_equations_cni} \; \\
    \frac{\mathrm{d} S_{m,l}}{\mathrm{d} t} =& \; \tilde{\alpha}_{m,l} I_{m,l} - \tilde{\beta}_{m,l} S_{m,l} \notag \\
                                            & + \theta_S \left [ (m-l + 1) S_{m,l-1} - (m-l) S_{m,l} \right ] \label{eq:master_equations_sml} \\
                                            & + \phi_S \left [ (l + 1) S_{m,l+1} - l S_{m,l} \right ] \;, \notag \\
    \frac{\mathrm{d} I_{m,l}}{\mathrm{d} t} =& -\tilde{\alpha}_{m,l} I_{m,l} + \tilde{\beta}_{m,l} S_{m,l} \notag \\
                                            & + \theta_I \left [ (m-l + 1) I_{m,l-1} - (m-l) I_{m,l} \right ] \label{eq:master_equations_iml} \\
                                            & + \phi_I \left [ (l + 1) I_{m,l+1} - l I_{m,l} \right ] \;. \notag
\end{align}
\end{subequations}
The four mean fields are calculated as
\begin{subequations}
\label{eq:mf}
\begin{align}
    \theta_S &= \frac{\sum_{n} (n-\bar{i}+1) (n-\bar{i}) C_{n,\bar{i}-1} \bar{\beta}_{n,\bar{i}-1}}{\sum_{n,i \leqslant \bar{i}-1} (n-i) C_{n,i}} \;, \label{eq:mf_th_s}\\
    \phi_S &= \frac{\sum_{n} (n-\bar{i}) \bar{i} C_{n,\bar{i}} \bar{\alpha}_{n,\bar{i}}}{\sum_{n,i>\bar{i}-1} (n-i) C_{n,i}} \;, \label{eq:mf_ph_s}\\
    \theta_I &= \frac{\sum_{n} \bar{i} (n-\bar{i}) C_{n,\bar{i}} \bar{\beta}_{n,\bar{i}}}{\sum_{n,i \leqslant \bar{i}} i C_{n,i}} \;, \label{eq:mf_th_i}\\
    \phi_I &= \frac{\sum_{n} (\bar{i}+1) \bar{i} C_{n,\bar{i}+1} \bar{\alpha}_{n,\bar{i}+1}}{\sum_{n,i>\bar{i}} i C_{n,i}} \;. \label{eq:mf_ph_i}
\end{align}
\end{subequations}
These quantities are the average rates at which inactive groups become active ($\theta_X$) or vice versa ($\phi_X$) given that we know the state ($X\in\{S,I\}$) of one of their members. We therefore sum over all groups eligible for the transition (e.g., with one too few or too many infectious nodes) and count the number of nodes therein whose state matches that of the node of interest. This gives us a biased distribution over states, renormalized with the sum in the denominator, and over which we average the local rate of transition. For instance, consider $\theta_S$ in Eq.~(\ref{eq:mf_th_s}), which is the average rate at which a random inactive group becomes active from the perspective of one of its susceptible members---the focal node. Since the probability that a group includes a susceptible node is proportional to the number of such nodes in the group (which is $n-i$ if the group has size $n$), the probability that an inactive group containing the susceptible node has size $n$ and $i = \bar{i}-1$ (i.e., is at the edge of becoming active) is $(n-\bar{i}+1) C_{n,\bar{i}-1}/\sum_{n,i \leqslant \bar{i}-1} (n-i) C_{n,i}$. Such a group has $n-\bar{i}$ susceptible nodes (except the focal node) and each of them gets infected with rate $\bar{\beta}_{n,\bar{i}-1}$, thus the local infection rate reads $(n-\bar{i}) \bar{\beta}_{n,\bar{i}-1}$. Summing over group sizes, we get Eq.~(\ref{eq:mf_th_s}). Equations~(\ref{eq:mf}b)-(\ref{eq:mf}d) are found analogously.

\begin{figure}
    \centering
    \includegraphics[width=0.9\linewidth]{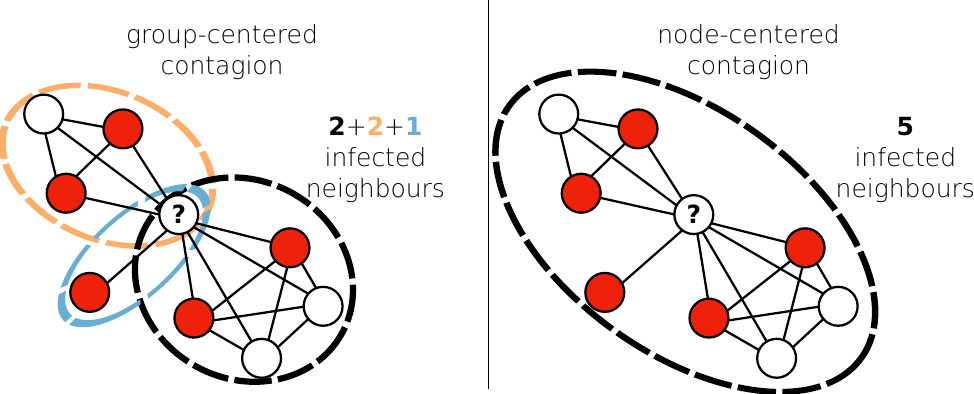}
    \caption{Conceptual difference between group-centered and node-centered dynamics. In the former, the contagion mechanism depends on how infected neighbors are distributed among the groups (depicted as cliques) of a focal node of interest. In the latter, this distribution does not matter and contagion is driven solely by the total number of infected neighbors.}
    \label{fig:mechanisms}
\end{figure}

To close the GAME, we need to estimate the previously introduced effective transition rates, $\bar{\alpha}_{n,i}$, $\bar{\beta}_{n,i}$, $\tilde{\alpha}_{m,l}$ and $\tilde{\beta}_{m,l}$. We calculate these rates with mean-field arguments, yet the form of this calculation depends on the nature of the dynamics. Specifically, we use two approaches based on whether the dynamics operate at the node or group level. In a \emph{node-centered} dynamics, the transition rates for a node are functions of the states of all of its neighbors, independently from how those neighbors are scattered across different groups. In a \emph{group-centered} dynamics, instead, each group brings an independent contribution so that the total transition rate is the sum of the transition rates over all the groups a node belongs to. In other words, node-centered and group-centered dynamics model mechanisms which are potentially nonlinear on the state of entire neighborhoods and single groups, respectively. The conceptual difference between the two mechanisms is illustrated in Fig.~\ref{fig:mechanisms}. As we prove in Methods, the two approaches become equivalent for linear dynamics (e.g., simple contagions).

To notice that Eqs.~(\ref{eq:master_equations}) and (\ref{eq:mf}) as well as all those derived in the two subsections below, are apt to describe any binary-state dynamics. The terms `susceptible' and `infected' carry no specific meaning at present---they can be mapped to spin up/down, adopter/non-adopter, cooperator/defector, etc. The dynamics of interest needs to be specified only after.

\subsubsection*{Node-centered dynamics}
\label{sec:node}

We first consider general continuous-time Markov processes where susceptible nodes become infected at rate $\beta(k,\ell)$, being $k$ the (pairwise) degree of the node and $\ell \in \lbrace 0, \dots, k \rbrace$ its infected degree. Similarly, infected nodes become susceptible at rate $\alpha(k,\ell)$. Degree $k$ and infected degree $\ell$ are total quantities of a node, summed over all groups it belongs to. The dynamics ignores how these quantities are distributed across groups and, therefore, can be seen as acting on the pairwise projection of the original higher-order network. Within this general node-centered process, we can calculate the mean-field transition rates as follows.

Let us first consider the effective infection rate $\bar{\beta}_{n,i}$. If we pick a susceptible node in a (focal) group of size $n$, of which $i$ are infected, the degree of this node can be decomposed as $k = n - 1 + r$, where $r$ is the excess degree, due to memberships to other groups. Similarly, the infected degree can be decomposed as $\ell = i + s$, where $s$ is the excess infected degree. If $r$ and $s$ were specified, then the infection rate of this susceptible node would simply be $\beta(n-1+r,i+s)$. In this version of the GAME, we approximate $\bar{\beta}_{n,i}$ by averaging over a joint distribution
\begin{align}
    \bar{\beta}_{n,i} = \sum_{r,s} \beta(n-1+r,i+s) P(r,s\vert n,i,S) \ ,
    \label{eq:beta_ni}
\end{align}
where the distribution $P(r,s\vert n,i,S)$ is to be determined. It is the probability that a susceptible node in a group of size $n$ with $i$ infected nodes has excess degree $r$ and infected excess degree $s$. For this task, we leverage the properties of probability generating functions (PGFs). Specifically, we need an expression for the following (bivariate) PGF,
\begin{align}
    E_S^i(x,y) = \sum_{r,s} P(r,s\vert n,i,S) x^r y^s \ ,
    \label{eq:E_S}
\end{align}
in terms of the state variables $\{C_{n,i}\}$ and $\{S_{m,l}\}$. Known $E_S^i(x,y)$, we can then extract the distribution $P(r,s\vert n,i,S)$ from Eq.~(\ref{eq:E_S}) using a discrete Fourier transform (see SM) and compute the effective rate $\bar{\beta}_{n,i}$ from Eq.~(\ref{eq:beta_ni}). To highlight that the latter is obtained averaging over the distribution associated with $E_S^i(x,y)$, we write
\begin{align}
    \bar{\beta}_{n,i} \equiv \langle \beta(n-1+r,i+s) \rangle_{E_S^i} \ .
\end{align}

If we take a random \textit{external} group to which the susceptible node belongs, the contribution to its degree and infected degree is associated with two PGFs, depending on whether or not the group is active for the node. If it is not (i.e., the group contains less than $\bar{i}$ infected nodes), then the appropriate PGF to use is
\begin{align}
    K_S^{i<\bar{i}}(x,y) &= \frac{\sum_{n,i<\bar{i}} (n-i) C_{n,i} x^{n-1} y^{i}}{\sum_{n,i<\bar{i}} (n-i) C_{n,i}} \ ;
    \label{eq:K_S_inact}
\end{align}
otherwise, it is
\begin{align}
    K_S^{i \geqslant \bar{i}}(x,y) &= \frac{\sum_{n,i \geqslant \bar{i}} (n-i) C_{n,i} x^{n-1}y^{i}}{\sum_{n,i \geqslant \bar{i}} (n-i) C_{n,i}} \ .
    \label{eq:K_S_act}
\end{align}
Through the auxiliary variables $x$ and $y$, these PGFs count the number of other members ($n-1$) and infected members ($i$) in a group, respectively, weighting these numbers with respect to the probability (proportional to $(n-i)C_{n,i}$) that a susceptible node is part of such a group. 

Equations~(\ref{eq:K_S_inact}) and (\ref{eq:K_S_act}) provide information about a single, random external group. But how many groups does the node belong to? Its excess membership and excess active membership are still unspecified. To compute them, we leverage again the information about whether the focal group is inactive ($i < \bar{i}$) or active ($i \geqslant \bar{i}$). Depending on this, a PGF of different form is used, i.e.,
\begin{align}
    G_S^i(x,y) &= \begin{dcases}
                \frac{\sum_{m,l} (m-l) S_{m,l} x^{m-l-1} y^{l}}{\sum_{m,l} (m-l) S_{m,l}} & \text{if } i < \bar{i}\ , \\
                \frac{\sum_{m,l} l S_{m,l} x^{m-l} y^{l-1}}{\sum_{m,l} l S_{m,l}} & \text{if } i \geqslant \bar{i}\ ,
             \end{dcases}
\end{align}
where the variables $x$ and $y$ here count the number of inactive and active external groups, respectively. If the focal group is inactive, then the probability for the susceptible node to be part of it is proportional to $(m-l) S_{m,l}$; if active, such probability is instead proportional to $l S_{m,l}$. 

Finally, assuming that the contributions to $r$ and $s$ from different groups are statistically independent 
(besides being identically distributed according to $K_S^{i<\bar{i}}$ or $K_S^{i \geqslant \bar{i}}$), using the properties of PGFs, it follows that $E_S^i(x,y)$ can be expressed through the composition
\begin{align}
    E_S^i(x,y) = G_S^i\left( K_S^{i<\bar{i}}(x,y), K_S^{i \geqslant \bar{i}}(x,y) \right) \ .
    \label{eq:E_S_explicit}
\end{align}

Analogous steps yield the effective recovery rate $\bar{\alpha}_{n,i}$ for an infected node, leading to
\begin{align}
    K_I^{i \leqslant \bar{i}}(x,y) &= \frac{\sum_{n,i\leqslant \bar{i}} i C_{n,i} x^{n-1} y^{i-1}}{\sum_{n,i \leqslant \bar{i}} i C_{n,i}} \ , \\
    K_I^{i > \bar{i}}(x,y) &= \frac{\sum_{n,i>\bar{i}} i C_{n,i} x^{n-1}y^{i-1}}{\sum_{n,i>\bar{i}} i C_{n,i}}\ , \\
    G_{I}^i(x,y) &= \begin{dcases}
                \frac{\sum_{m,l} (m-l) I_{m,l} x^{m-l-1} y^{l}}{\sum_{m,l} (m-l) I_{m,l}} & \text{if } i \leqslant \bar{i}\ , \\
                \frac{\sum_{m,l} l I_{m,l} x^{m-l} y^{l-1}}{\sum_{m,l} l I_{m,l}} & \text{if } i > \bar{i}\ ,
             \end{dcases} \\
    E_I^i(x,y) &= G_I^i\left( K_I^{i \leqslant \bar{i}}(x,y), K_I^{i>\bar{i}}(x,y) \right) \ ,
\end{align}
and eventually to
\begin{align}
    \bar{\alpha}_{n,i} \equiv \langle \alpha(n-1+r,i-1+s) \rangle_{E_I^i} \ .
\end{align}
Recall that, from the perspective of an infected node, a group is active if there are at least $\bar{i}$ infected nodes among the other members.

The construction of the PGFs to compute $\tilde{\alpha}_{m,l}$ and $\tilde{\beta}_{m,l}$ is straightforward, for the memberships are already given. For an infected node with membership $m$ and $l$ of them active, we need the distribution $P(k,\ell \vert m,l,I)$ for its degree $k$ and infected degree $\ell$. The associated PGF reads
\begin{align}
    \notag E_I^{m,l}(x,y) =& \sum_{k,\ell} P(k,\ell \vert m,l,I) x^k y^\ell \\ =& \left [ K_I^{i \leqslant \bar{i}}(x,y)\right ]^{m-l} \left [ K_I^{i>\bar{i}}(x,y)\right ]^l \ ,
\end{align}
where, for the second equality, we assumed that the contributions to $k$ and $\ell$ from different groups are independent. The effective recovery rate is then
\begin{align}
    \tilde{\alpha}_{m,l} \equiv \langle \alpha(k,\ell) \rangle_{E_I^{m,l}} \ .
\end{align}
Analogously, for a susceptible node, we have
\begin{align}
    \notag E_S^{m,l}(x,y) =& \sum_{k,\ell} P(k,\ell \vert m,l,S) x^k y^\ell \\ =& \left [ K_S^{i<\bar{i}}(x,y)\right ]^{m-l} \left [ K_S^{i \geqslant \bar{i}}(x,y)\right ]^l \ ,
\end{align}
and the effective infection rate reads
\begin{align}
    \tilde{\beta}_{m,l} \equiv \langle \beta(k,\ell) \rangle_{E_S^{m,l}} \ .
\end{align}
This closes the GAME for node-centered dynamics. Once a PGF  has been computed, we can extract the probability distribution it generates as detailed in Methods. 

\begin{figure*}
    \centering
    \includegraphics[width=\linewidth]{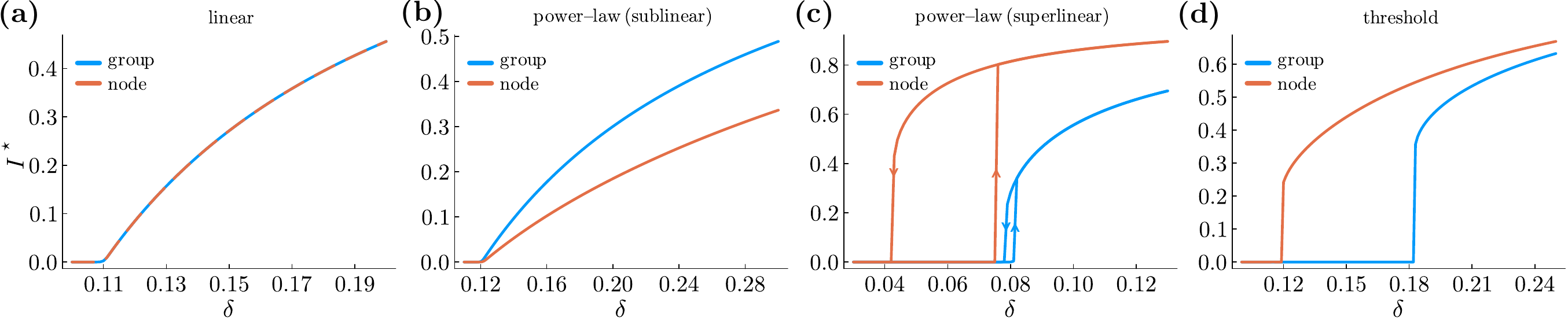}
    \caption{Comparison of the equilibrium prevalence, $I^\star$, obtained integrating Eqs.~(\ref{eq:master_equations}) on random 3-regular hypergraphs under node-centered (blue curves) and group-centered dynamics (red curves), considering four different forms for the infection kernels. The group size distribution is a truncated Poisson with mode $n=4$ and support $\{2,\dots,6\}$ in (a-c) and with mode $n=5$ and support $\{3,\dots,7\}$ in (d). (a) Linear kernels, $\beta(k,\ell) = \delta \ell$ (node) and $\lambda(n,i) = \delta i$ (group); $I(0)=0.01$ and $\bar{i} = 1$. Node- and group-centered dynamics are equivalent in this case (see SM for analytical proof). (b) Sublinear kernels, $\beta(k,\ell) = \delta \ell^{1/2}$ (node) and $\lambda(n,i) = \delta i^{1/2}$ (group); $I(0)=0.01$ and $\bar{i} = 1$. (c) Superlinear kernels, $\beta(k,\ell) = \delta \ell^{2}$ (node) and $\lambda(n,i) = \delta i^{2}$ (group); $I(0)=0.01$ for lower branches and $I(0)=0.8$ for upper branches, and $\bar{i} = 1$. (d) Threshold kernels, $\beta(k,\ell) = {\bf 1}_{\ell \geqslant 2} \delta \ell$ (node) and $\lambda(n,i) = {\bf 1}_{i \geqslant 2} \delta i$ (group); $I(0)=0.8$ and $\bar{i} = 2$.}
    \label{fig:nodeVSgroup}
\end{figure*}

\subsubsection*{Group-centered dynamics}
\label{sec:group}

We now consider the case where groups are the main actors responsible for transitions. Specifically, a susceptible node in a group of size $n$ with $i$ infectious members receives an infection rate of $\lambda(n,i)$ from this group. The overall, effective infection rate $\bar{\beta}_{n,i}$ thus becomes $\lambda(n,i) + \bar{\lambda}_i$, where $\bar{\lambda}_i$ is the average infection rate from all the external groups, reading
\begin{align}
    \notag \bar{\lambda}_i &= \begin{dcases}
                        		 	\frac{\sum_{m,l} (m-l) S_{m,l} \left[ (m\!-\!l\!-\! 1) \bar{\lambda}_{i<\bar{i}} + l \bar{\lambda}_{i \geqslant \bar{i}} \right]}{\sum_{m,l} (m-l) S_{m,l} } & \text{if } i < \bar{i}
                        		 	\\
                        		 	\frac{\sum_{m,l} l S_{m,l} \left[ (m-l) \bar{\lambda}_{i<\bar{i}} + (l-1) \bar{\lambda}_{i \geqslant \bar{i}} \right]}{\sum_{m,l} l S_{m,l} } & \text{if } i \geqslant \bar{i}
                    		 	\end{dcases} \\
		 	    	      &= \left[\bar{\lambda}_{i<\bar{i}}\frac{\partial}{\partial x} + \bar{\lambda}_{i \geqslant \bar{i}}\frac{\partial}{\partial y}\right]_{x,y=1} G_S^i(x,y) \ .
    \label{eq:lambda_i}
\end{align}
The quantities $\bar{\lambda}_{i < \bar{i}}$ and $\bar{\lambda}_{i \geqslant \bar{i}}$ are the average infection rates from a random external inactive or active group (for a susceptible node), respectively, and read
\begin{align}
    \bar{\lambda}_{i < \bar{i}} &= \frac{\sum_{n,i < \bar{i}} (n-i) C_{n,i} \lambda(n,i)}{\sum_{n,i < \bar{i}} (n-i) C_{n,i}} = \langle \lambda(n,i) \rangle_{K_S^{i<\bar{i}}}
    \ , \label{eq:lambda_ilj} \\
    \bar{\lambda}_{i \geqslant \bar{i}} &= \frac{\sum_{n,i \geqslant \bar{i}} (n-i) C_{n,i} \lambda(n,i)}{\sum_{n,i \geqslant \bar{i}} (n-i) C_{n,i}} = \langle \lambda(n,i) \rangle_{K_S^{i \geqslant \bar{i}}}
    \label{eq:lambda_igj} \ .
\end{align}
From Eq.~(\ref{eq:lambda_i}) we see that $\bar{\lambda}_i$ results from the sum of two terms, each one given by the product between the average infection rate in an external group of a given activity state and the average number of such groups a susceptible node belongs to. Notice that, although the rates $\bar{\lambda}_i$, $\bar{\lambda}_{i < \bar{i}}$ and $\bar{\lambda}_{i \geqslant \bar{i}}$ can be expressed in terms of PGFs, we do not need the latter to compute the rates, for these are just sums over state variables. The reason for this simplification is that the overall effect on a node is just a linear combination of the effects produced by each group the node belongs to, whereas the effects coming from different neighbors could be combined nonlinearly in the node-centered dynamics.

Next, the effective rate $\tilde{\beta}_{m,l}$ simply reads
\begin{equation}
     \tilde{\beta}_{m,l} = (m-l) \bar{\lambda}_{i < \bar{i}} + l \bar{\lambda}_{i \geqslant \bar{i}}  \ ,
     \label{eq:group_beta_ml} 
\end{equation}
as we have full knowledge of the memberships $m$ and $l$.

The recovery rates $\bar{\alpha}_{n,i}$ and $\tilde{\alpha}_{m,l}$ are computed analogously by estimating the average recovery rates $\bar{\mu}_{i \leqslant \bar{i}}$ and $\bar{\mu}_{i > \bar{i}}$ of infected nodes in inactive and active groups, respectively, given the within-group recovery rate $\mu(n,i)$.

\subsection*{Results on static structures}\label{sec:results_game}

\subsubsection*{Node-centered versus group-centered dynamics}

We start our exploration of the model by comparing node- and group-centered dynamics using different functional forms for the infection kernel. Throughout these examples, recovery is considered as a spontaneous node transition occurring at constant rate $1$, implying  $\bar{\alpha}_{n,i} = \tilde{\alpha}_{m,l} = 1$. While the two dynamics are equivalent for linear kernels (see Supplementary Material), used for instance to model simple contagions, they produce different outcomes when kernels depend nonlinearly on infectious contacts, as is the case for complex contagions~\cite{PhysRevLett.92.218701, Guilbeault2018}. We show this using power-law kernels of the form $\delta i^\sigma$ with $\sigma < 1$ (sublinear) or $\sigma > 1$ (superlinear), as well as threshold kernels of the form ${\bf 1}_{i \geqslant \nu} \delta i$, being zero for $i < \nu$ and linear for $i \geqslant \nu$.
Power-law kernels are postulated by social impact theory~\cite{latane1981psychology} and offer more suited descriptions when the linear assumption breaks down, for example, in the presence of saturation effects or when exposure to multiple infections becomes important~\cite{liu1987dynamical}. On the other hand, the threshold kernel above turns out to be a good effective model to approximate contagions unfolding in uncertain transmission settings~\cite{st2024nonlinear}.
We illustrate the difference between node- and group-centered dynamics under the aforementioned kernels in Fig.~\ref{fig:nodeVSgroup}, depicting the equilibrium prevalence, $I^\star$, against $\delta$. Notice that, here and after, Eqs.~(\ref{eq:master_equations}) are initialized by distributing the proportion $I(0)$ of infections uniformly at random, which means that $C_{n,i}(0) = p_n\binom{n}{i}I(0)^i (1-I(0))^{n-i}$, $I_{m,l}(0)=I(0) g_m \binom{m}{l} Q^l (1-Q)^{m-l}$, and $S_{m,l}(0)=(1-I(0)) g_m \binom{m}{l} Q^l (1-Q)^{m-l}$; being $Q = \sum_n n p_n \sum_{i\geqslant\bar{i}}\binom{n-1}{i}I(0)^i (1-I(0))^{n-1-i}$ the probability that a random group is active.

The node-centered dynamics favors the contagion by increasing $I^\star$ and decreasing the critical thresholds compared to the group-centered dynamics, for both superlinear and threshold infection kernels. The opposite holds under sublinear kernels, although the effect on the invasion threshold is tiny in this case (imperceptible for the magnitudes shown in Fig.~\ref{fig:nodeVSgroup}(b)). The difference in the outcomes of the two dynamics is easily understood from the functional forms of the kernels.
Let $i = \sum_{k=1}^m i_k$ be the total number of infected nodes in the neighborhood of a susceptible node, being $\{i_k\}_{k=1}^m$ the configuration of the number of infected nodes in each of the $m$ groups the susceptible node belongs to.
Then, for power-law kernels,  $i^\sigma \lessgtr \sum_{k=1}^m i_k^\sigma$ if $\sigma \lessgtr 1$. Similarly, ${\bf 1}_{i \geqslant \nu} i \geqslant \sum_{k=1}^m {\bf 1}_{i_k \geqslant \nu} i_k$ for threshold kernels, where the equality holds only if $i_k \geqslant \nu$ for all $k$.

\begin{figure}
    \centering
    \includegraphics[width=\linewidth]{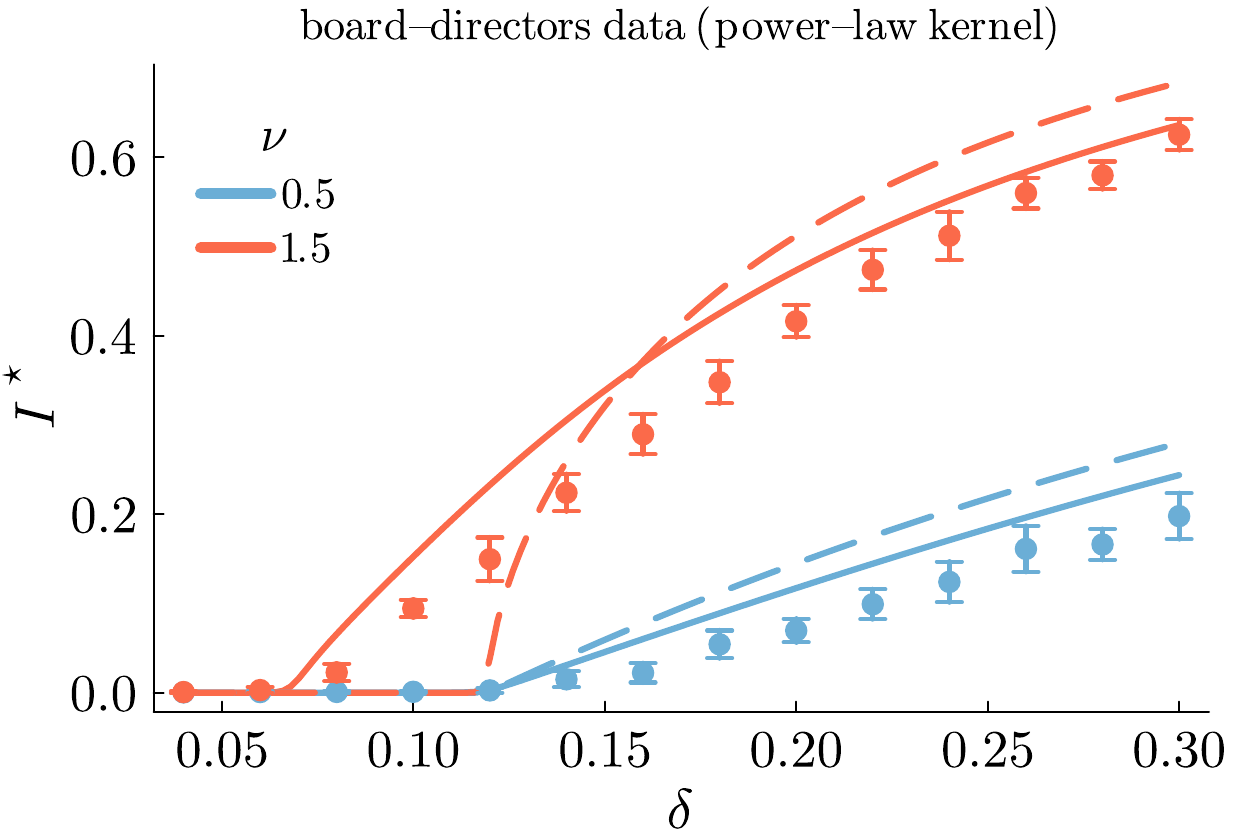}
    \caption{Results on a hypergraph built from data about board directors co-sitting on common boards~\cite{seierstad2011few,netzschleuder}. Equilibrium prevalence, $I^\star$, under group-centered dynamics considering sublinear ($\nu = 0.5$) and superlinear ($\nu = 1.5$) infection kernels $\lambda(n,i) = \delta i^{\nu}$; $I(0)=0.8$. Solid and dashed lines represent the results obtained integrating the GAME (Eqs.~(\ref{eq:master_equations}) for $\bar{i}=1$) and a heterogeneous mean-field approximation (see SM), respectively. Points and error bars denote means and standard errors over $20$ random Monte Carlo realizations.}
    \label{fig:real_world}
\end{figure}

\begin{figure*}
    \centering
    \includegraphics[width=\linewidth]{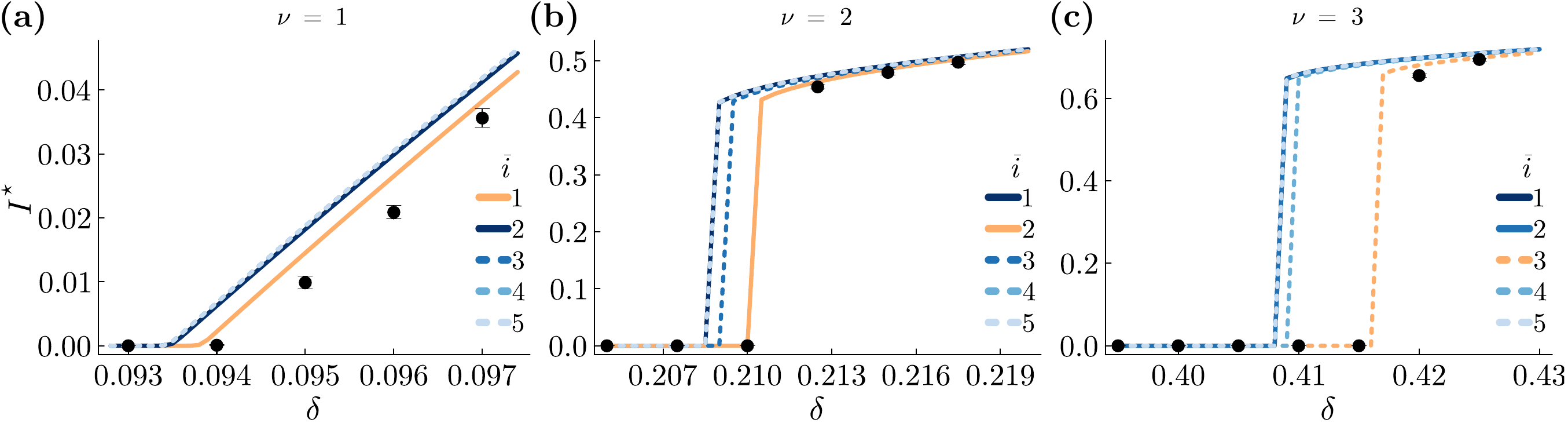}
    \caption{Equilibrium prevalence, $I^\star$, obtained on random 3-regular 5-uniform hypergraphs under group-centered dynamics considering a threshold infection kernel $\lambda(n,i) = {\bf 1}_{i \geqslant \nu} \delta i$, for (a) $\nu=1$ ($I(0)=0.06$), (b) $\nu=2$ ($I(0)=0.8$) and (c) $\nu=3$ ($I(0)=0.8$). Solid and dashed lines represent the results obtained integrating Eqs.~(\ref{eq:master_equations}) for $\bar{i}\in\{1,\dots,5\}$, while points and error bars (when visible) denote averages and standard errors over $20$ random realizations resulting from Monte Carlo simulations performed on hypergraphs with $N=5\times 10^4$ nodes. Notice that the model with $\bar{i} = 5$ is equivalent to the group-based AME~\cite{hebert2010propagation}. The most accurate model is always the one associated to $\bar{i}=\nu$ (orange curves).}
    \label{fig:threshold}
\end{figure*}

This difference can help us reinterpret known empirical results showing that some social contagions are promoted when exposures are scattered across multiple groups~\cite{ugander2012structural}. On one hand, such evidence does not support node-centered dynamics, this being agnostic to how contacts are distributed over groups. On the other hand, for superlinear or threshold mechanisms (e.g., peer pressure or social reinforcement)~\cite{centola2010spread}, our results imply that the dynamics cannot be group-centered either, for transmission would be maximized if all of a node's active contacts came from a single group. However, if the mechanisms at play were instead sublinear (e.g., saturation or hipster effects), the empirical findings would be compatible with a group-centered dynamics, as the latter maximizes transmission when the infectious contacts are distributed uniformly across a node's groups. In fact, Ugander \textit{et al.}~\cite{ugander2012structural} report an infection kernel that increases sublinearly to then eventually decrease. Combined with our results, this provides evidence for transmissions being group-centered.

To the best of our knowledge, the analysis in Ref.~\cite{ugander2012structural} is unique in its ability to account for how exposure is distributed across different groups. The conclusions above, therefore, cannot be generalized to all sorts of social contagion. Some contagions may spread via node-centered or group-centered transmissions, others via more involved dynamics. Understanding how different contagions work in a group-structured social setting thus necessitates new experimental and observational studies. In addition to making predictions, our GAME framework can be used to guide those empirical studies and, as shown above, to offer a mechanistic interpretation of their findings.

We tested the GAME by comparing their predictions to Monte Carlo simulations performed on structures built from real-world data (see the Supplemental Material (SM) for details). Figure~\ref{fig:real_world} reports the results obtained for a hypergraph generated from records of board directors (nodes) co-sitting on common boards (hyperedges)~\cite{seierstad2011few,netzschleuder}. Despite the limited size of the system (870 nodes) and, more importantly, a relevant fraction (nearly 25\%) of hyperedges that overlap over two or more nodes, the GAME provides good predictions overall. Notice, in fact, that the model implicitly assumes an infinite system and an asymptotically vanishing fraction of hyperedges overlapping over multiple nodes. To show how prediction performance is affected by neglecting local dynamical correlations, we also report the results obtained from a heterogeneous mean-field theory (HMF) that preserves the group size and membership distributions (see Methods for derivation). The HMF offers substantially poorer predictions and specifically ignores the dependence of the invasion threshold on nonlinear higher-order channels of infection. A reason for the decrease in accuracy is the inability to capture the localization of activity within groups of different sizes~\cite{hebert2010propagation,st-onge2021master,st-onge2022influential} and around nodes of different memberships (see SM for details). Nonetheless, even when localization is prevented by considering uniform and regular networks, HMF's accuracy can drastically drop because of structural sparsity (see SM for further results).

While in this section we focused on group-centered dynamics, in the SM we tested the performance of the GAME under node-centered dynamics. The model proves to be highly accurate even when node-based AME~\cite{marceau2010adaptive} become unreliable.

In what follows, the distinction between group- and node-centered transmission does not matter qualitatively. Being faster to integrate, we chose to rely on the former.

\subsubsection*{Dynamical correlations}

In the previous section, we tacitly set the activity threshold $\bar{i}$ of the model equal to either $1$ or $2$ depending on the functional form chosen for the infection kernel. The reason is that, as we show below, those are the values of $\bar{i}$ that most accurately reproduce the stochastic system simulated using those kernels.

Compared to previous approaches~\cite{hebert2010propagation,st-onge2021social,st-onge2021master}, the GAME model improves the description of the system by accounting for the activity around nodes, preserving in this way dynamical correlations across adjacent groups that would be thoroughly ignored otherwise. It does so by coarse-graining the activity configurations of groups around a node by tagging each group as \textit{active} or \textit{inactive} based on how its activity level (net of the contribution of the focal node) compares to $\bar{i}$. We therefore look for the value of $\bar{i}$ that, by tagging groups to capture the most relevant correlations, best approximates the stochastic process.

The basic expectation is that the best model is the one that, from the perspective of a susceptible node, labels as \textit{active} only the groups that \textit{can} transmit, i.e., with a nonzero infection kernel. We test this hypothesis on synthetic hypergraphs using a group-centered dynamics with threshold kernel $\lambda(n,i) = {\bf 1}_{i \geqslant \nu} \delta i$, setting the threshold for transmission to $\nu = 1,2,3$. In accordance with the GAME's assumption that the structure is random, the hypegraphs are generated using a bipartite configuration model. Concisely, we first draw $N$ nodes from $g_m$ and $M$ hyperedges (groups) from $p_n$, making sure the constraint $N\langle m\rangle=M\langle n\rangle$ holds, and then match node-stubs with group-stubs uniformly at random, until no free stubs are left. As Figure~\ref{fig:threshold} shows, when comparing with Monte Carlo simulations, the most accurate model is indeed the one with $\bar{i} = \nu$ (the result applies to node-centered dynamics too).

A relevant alternative form for the kernel could be a sigmoid with a maximal steepness at an inflection point $x_0$, this working as an effective threshold. Consider the case where the infection kernel stays positive but close to zero up to $i=x_0-1>0$, to then rapidly accelerate to a much larger value when approaching $i=x_0$, before eventually saturating. In the SM we show how in such case the best $\bar{i}$ coincides with $x_0$, even if the kernel is nonzero at $i=1$. The only exception to this regards the invasion threshold, which is better located by setting $\bar{i}=1$. This can be expected, for most infectious groups in a nearly susceptible population contain just one infected node.

Taken together, these findings corroborate the intuition that the best model is the one that identifies the most appropriate boundary separating groups able to self-sustain the spread from those unable to do so. For kernels with a strong scale such as steep sigmoids or sharp thresholds, our results demonstrate that the optimal $\bar{i}$ can be readily identified a priori. Likewise, in the absence of steep changes, as is the case for linear and power-law kernels like those used in Figs.~\ref{fig:nodeVSgroup}(a)-(c), the best $\bar{i}$ simply coincides with the minimum number of infected nodes required for transmission. In intermediate cases where the kernel shows a somewhat weak scale (e.g., a not-to-steep sigmoid), multiple values of $\bar{i}$ provide very close predictions, making the choice of $\bar{i}$ unimportant.

Lastly, in a regime of mesoscopic localization where the contagion is disproportionately concentrated within the largest groups~\cite{st-onge2021social,st-onge2021master}, one could expect that a higher $\bar{i}$ might be optimal because discerning large, highly active groups from small, lowly active ones. The results obtained using a linear kernel (see SM) demonstrate that the most appropriate activity threshold is still $\bar{i} = 1$, as in a delocalized regime (Fig.~\ref{fig:threshold}(a)). We do not exclude, however, that mesoscopic localization could select a different optimal $\bar{i}$ for heavy-tailed group size distributions, whose investigation is left for future work.

In summary, through the activity scale $\bar{i}$, the GAME model can be readily tailored to capture the most relevant dynamical correlations for the specific dynamics under study, allowing for more accurate predictions.




\subsection*{Results on adaptive structures}

The very same formalism developed in Methods provides a suitable framework to study group-structured \textit{adaptive} systems. Some knowledge of the (activity) state of the groups is indeed required to describe interesting adaptive responses~\cite{gross2009adaptive}; otherwise, one could only model na\"{i}ve agents leaving and joining groups entirely at random. The adaptive GAME (or A-GAME) presented below, therefore, opens new possibilities for the modeling of adaptive systems.

Consider, for example, susceptible/cooperative agents trying to avoid infectious/defective environments. Dynamics of this kind can be described with our formalism by reinterpreting the characteristic activity scale $\bar{i}$ as a tolerance threshold, such that agents try to rewire away from active groups. Of course, other meanings can be associated to $\bar{i}$; for instance, activity could be an attractive trait, so that agents would aim to connect to active groups.

\begin{figure*}
    \centering
    \includegraphics[width=\linewidth]{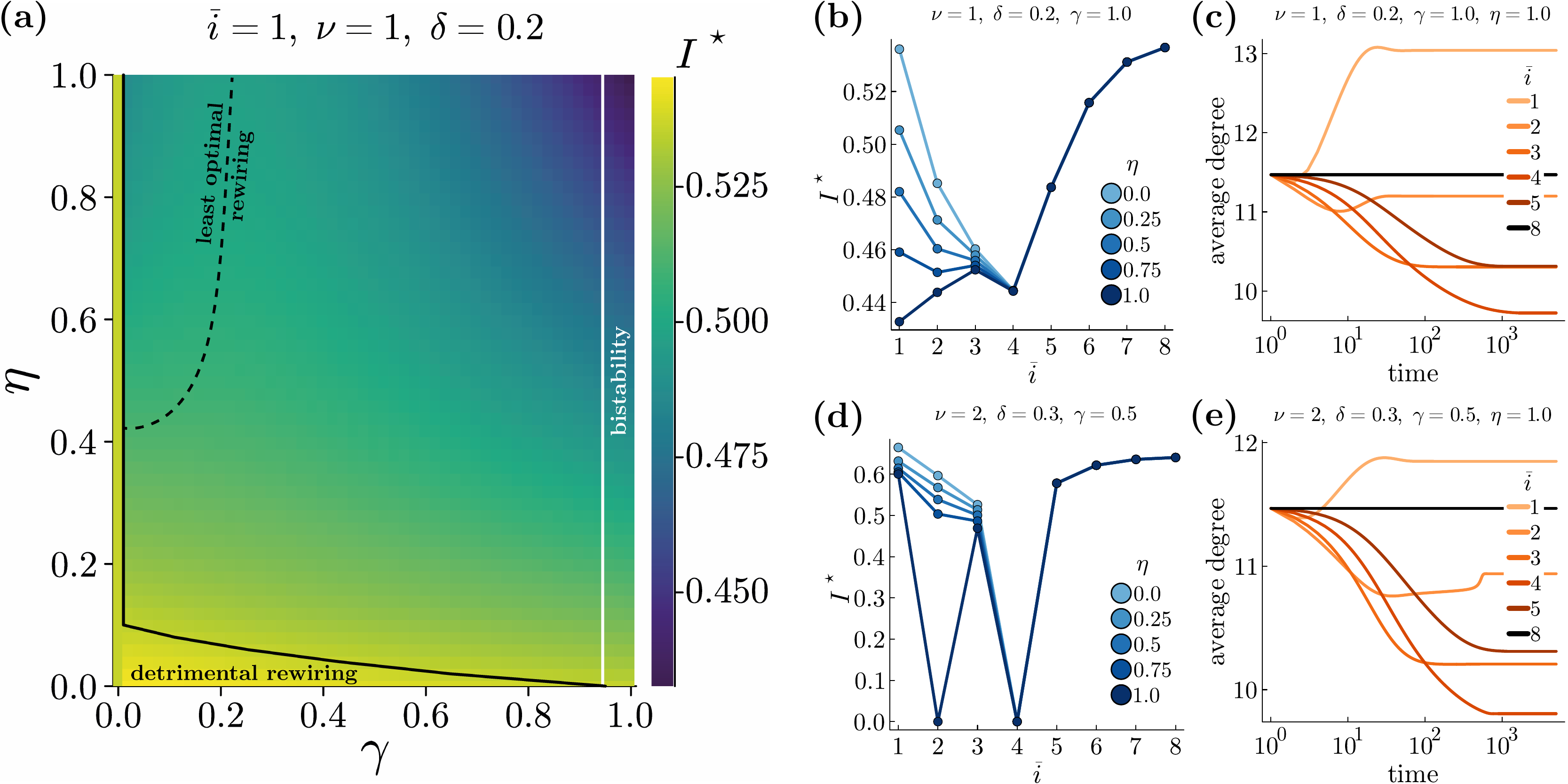}
    \caption{Results for adaptive hypergraphs (Eqs.~(\ref{eq:rewiring})). The initial size distribution is a truncated Poisson with mode $n=4$ and support $\{2,\dots,8\}$ ($n=8$ being a maximal group size), yielding an average group size $\langle n\rangle = 4.196$. All nodes have membership $m=3$.
    (a) Phase diagram for a simple contagion ($\nu = 1$) showing the equilibrium prevalence $I^\star$ against rewiring rate $\gamma$ and rewiring accuracy $\eta$, when strategy $\bar{i}=\nu=1$ is used. We observe two novel phenomena: (i) a low-accuracy region, on the left of the solid black curve, where rewiring is detrimental, for it leads to a higher prevalence than with no rewiring (notice that above $\eta \approx 0.1$ the region only consists of the bin corresponding to no rewiring, $\gamma = 0$); (ii) a slow-rewiring region where, for high enough accuracy, prevalence first increases and then decreases with $\gamma$, determining for each value of $\eta$  a least optimal rewiring rate (black dashed curve) at which prevalence is maximal, yet still lower than in the absence of rewiring. Additionally, by increasing the rewiring rate, we eventually encounter the $\eta$-independent invasion threshold (vertical white line) marking the onset of a bistable region (notice that the $\eta$-dependent persistence threshold lies outside the area here shown).
    (b) Varying $\bar{i}$ we find two main strategies for the agents to reduce (and possibly eradicate) contagions. At high $\gamma$ and $\eta$, it is optimal for agents to target the dynamics by setting $\bar{i}=\nu$, which is also how modelers can optimize their model as seen in Fig.~\ref{fig:threshold}. At low $\eta$ or low $\gamma$ (see SM for the latter case), it is optimal to target instead the structure and minimize degree by setting $\bar{i} = 4 \approx \langle n\rangle$.
    (c) Time evolution of the average degree in case (b) for $\eta = 1.0$. Strategies $\bar{i}=\nu$ and $\bar{i} = 4 \approx \langle n\rangle$ are the two best, but only the latter explicitly targets the structure in order to reduce the number of contacts. (d \& e) Analogous to panels (b \& c) but for complex contagion ($\nu = 2$).
    }
    \label{fig:adaptive}
\end{figure*}

Here we consider the case where only susceptible nodes can rewire away from active groups (the general model where both susceptible and infected agents can rewire~\cite{scarpino2016effect} is provided in the SM). To quantify the information nodes have about groups before joining them, we define the probability $\eta$ that rewiring is targeted towards inactive groups, as opposed to random groups in any state. To notice that, in this framework, the original adaptive network model by Gross \emph{et al.}~\cite{gross2006epidemic} becomes the special case where all groups are just pairs and susceptible agents have perfect information ($\eta = 1$).

With group rewiring, the average group size and average group membership are conserved, but the active membership of nodes and the size distribution of groups (and consequently both the degree distribution and the average degree of the projected network) are allowed to change as an adaptive response to the contagion. In principle, this model can interpolate between a complete network with a giant infinite group and a sparse regular network.
 
Rewiring adds the following transitions to the GAME,
\begin{subequations}
\label{eq:rewiring}
\begin{align}
    \notag \frac{\mathrm{d}C^{\textrm{a}}_{n,i}}{\mathrm{d}t} =& \; \gamma {\bf 1}_{i \geqslant \bar{i}} \left[(n+1-i)C_{n+1,i} - (n-i)C_{n,i}\right] \label{eq:rewiring_cni} \\
            &+ \gamma \Omega_{S\vert i\geqslant \bar{i}} \left( \frac{\eta {\bf 1}_{i < \bar{i}}}{C_{i<\bar{i}}} + 1-\eta \right) \left(C_{n-1,i} - C_{n,i} \right) \;, \\
    \frac{\mathrm{d} S^{\textrm{a}}_{m,l}}{\mathrm{d} t} =& \; \gamma \left[\eta + (1-\eta) C_{i<\bar{i}}\right] \left[(l+1)S_{m,l+1}-lS_{m,l}\right] \;. \label{eq:rewiring_sml}
\end{align}
\end{subequations}
These rates of change are tagged with `${\textrm{a}}$' for `adaptive' and added to Eqs.~(\ref{eq:master_equations_cni}) and (\ref{eq:master_equations_sml}), respectively (see SM for the full equations).
The two terms in Eq.~(\ref{eq:rewiring_cni}) accounts for susceptible nodes leaving and joining groups, respectively, while Eq.~(\ref{eq:rewiring_sml}) only needs to account for susceptible nodes rewiring away from their active groups. We defined $C_{i<\bar{i}} = \sum_{n,i<\bar{i}} C_{n,i}$ and  $\Omega_{S \vert i \geqslant \bar{i}} = \sum_{n,i\geqslant \bar{i}} (n-i) C_{n,i}$.

We experiment with adaptive hypergraphs considering group-centered dynamics in Fig.~\ref{fig:adaptive}. We use threshold infection kernels of the form $\lambda(n,i) = {\bf 1}_{i \geqslant \nu} \delta i$ to explore both simple ($\nu=1$) and complex threshold-like contagions ($\nu=2$). The phase diagram of the dynamics highlights a few important results. For simple contagion, there exists a bistable region for any rewiring accuracy when adaptation is fast enough, generalizing the results from Gross \emph{et al}.~\cite{gross2006epidemic}. As we show in the SM, when described as a function of the infection rate $\delta$, the bistable region is widened by either increasing the rewiring rate---the invasion threshold increases faster than the persistence one---or decreasing the rewiring accuracy---the invasion threshold is unaffected while the persistence one decreases. For complex contagion, the system shows bistability as already does in the static case---rewiring moves the persistence threshold but does not produce new equilibria.

More importantly, the higher-order organization leads to two previously unseen phenomena (see Fig.~\ref{fig:adaptive}(a)). First, a low-accuracy region of \emph{detrimental rewiring}, where prevalence is higher than with no rewiring. Second, a slow-rewiring region where, for high enough accuracy, prevalence is lower than with no rewiring, but increases with $\gamma$ before eventually decreasing, defining a value of \emph{least optimal rewiring rate} for each $\eta$.

To better understand the rich phenomenology of higher-order adaptation, we identify two strategies for agents in the network to escape the contagion (see Figs.~\ref{fig:adaptive}(b)-(e)):
\begin{enumerate}
    \item \emph{Avoid contagious groups}. This is optimal when targeting is both fast (high $\gamma$) and accurate (high $\eta$). To do so, nodes have to mimic modelers and set $\bar{i}=\nu$. Being enough reactive and precise, they can manage to escape infection, even without either lowering or minimizing the connectivity of the structure; 
    \item \emph{Avoid large groups}. This is optimal when rewiring is either slow (low $\gamma$) or 
 poor (low $\eta$). By setting $\bar{i} \approx \langle n \rangle$, the average group size, nodes rewire away from groups that are larger than average, thereby minimizing their average degree and the probability of getting infected.
\end{enumerate}\

It is interesting that the most direct and intuitive adaptive strategy---trying to leave a group as soon as it becomes contagious (i.e., $\bar{i} = \nu$)---is actually optimal only in a tiny region of the $\gamma$-$\eta$ space, performing generally poorly elsewhere. The complexity of the process prevents us however from providing a clear, quantitative boundary in the parameter space defining which strategy is optimal in each situation.
In fact, in intermediate regimes, both strategies can work just as well but an intermediary strategy ($\nu < \bar{i} < \langle n \rangle$) may not, and is never optimal anyway (see Figs.~\ref{fig:adaptive}(b) and (d)). We hypothesize that this is because the mechanisms underlying these strategies are actually in opposition.

To look at the mechanisms explaining these results, we show the temporal density of the adaptive hypergraphs under different rewiring strategies in Figs.~\ref{fig:adaptive}(c) and (e). We see that the $\bar{i}=\nu$ strategy works despite slightly decreasing ($\nu=2$) or even increasing ($\nu=1$) the connectivity of the system.
As expected, the $\bar{i}\approx\langle n \rangle$ strategy works by having susceptible agents avoid groups larger than average and therefore creates a more uniform sparse network, consequently minimizing the average degree. Indeed, the latter is equal to $\langle m \rangle \langle n(n-1) \rangle / \langle n \rangle$, and being the average membership $\langle m \rangle$ and the average group size $\langle n \rangle$ conserved quantities, it is simply proportional to $\langle n^2 \rangle$ (or, equivalently, to the group size variance, $\langle n^2 \rangle - \langle n \rangle^2$). In accordance to our hypothesis, the two adaptive strategies work in different ways, targeting either dynamics---contagious groups---or structure---large groups.

A similar logic explains the observed region of least optimal rewiring rates. Larger groups reach $i = \bar{i}$ faster on average and slow rewiring allows $i$ to significantly correlate with group size before the typical rewiring time ($1/\gamma$). For high enough accuracy, susceptible agents then preferentially migrate to smaller groups, decreasing the average degree (see SM). Conversely, fast rewiring makes targeting the dynamics the optimal strategy. In between, we find a least optimal rewiring rate that is too slow to avoid the dynamics but too fast to minimize the degree.

\section*{Discussion}

We studied contagions on static and adaptive hypergraphs by developing a general model able to capture both intra- and inter-group dynamical correlations.
To do so, we introduced the notion of characteristic scale $\bar{i}$ of a contagion to tag groups as active or inactive based on the number of infectious nodes they contain. Our GAME thus parsimoniously generalize node- and group-based AME approaches, which are respectively recovered when one collapses groups to pairs and considers all groups as equivalently (in)active. Hence, whether the aim is to describe binary-state dynamics on networks or hypergraphs, either static or adaptive, it's in the GAME.

We asked three related questions. How does a hypergraph contagion depends on the scale at which its transmission mechanism operates? What is the characteristic scale $\bar{i}$ at which our model can best capture the dynamics on static hypergraphs? Allowing agents to adaptively rewire their memberships, what is the characteristic scale $\bar{i}$ that allows them to best avoid the contagion?

We first demonstrated that group-centered transmission mechanisms help sublinear contagions spread and may explain empirical results on the importance of having contagious contacts distributed across multiple groups. 
Regarding the characteristic scale $\bar{i}$, we found that modelers should use the value of $\bar{i}$ that best discerns sufficiently from insufficiently contagious groups---a separation generally deducible a priori by simply looking at the functional form of the infection kernel.
In adaptive hypergraphs, agents instead have multiple suitable options to avoid the contagion. In the exceptional situation in which rewiring is both very fast and accurate in targeting non-infectious groups, agents should use the same value of $\bar{i}$ modelers would use, thus reducing the contagion events without needing to minimize (or even lower) the overall connectivity. However, whenever rewiring is either slow or inaccurate, agents should instead use $\bar{i}\approx\langle n \rangle$ to minimize their degree by rewiring away from groups larger than average.

Importantly, the GAME allowed us to introduce and start exploring adaptive hypergraphs as it captures correlations both within and across groups. Adaptive hypergraphs are not as constrained as most adaptive network models are, for their density or average degree is not fixed over time, and can thus self-organize in diverse ways. 
Even with fixed average membership (hyperdegree), the A-GAME can track very sparse as well as very dense networks as groups sizes fluctuate.
By covering the space between these two limits, adaptive hypergraphs can help explore the richness of network structures produced by adaptive mechanisms in nature.

In human populations, groups have long been recognized as a vehicle for cultural dynamics \cite{cavalli1981cultural, boyd1988culture}. Groups can often outlive their members and, through shared norms or behaviors, groups can effectively act as agents in the social dynamics just like individuals do. 
For instance, a famous case study in network science, Zachary’s Karate Club \cite{zachary1977information}, exemplifies group conflict, fission, and adaptation. The research describes a social club where tensions due to asymmetric flow of information lead to the formation of new subgroups---however, it has been mostly studied through the lens of static pairwise networks so far.
The rise of mathematical models for hypergraphs thus provides a unique opportunity to jointly model the complex dynamics of individuals and of the groups they compose. In fact, recent work has explored this exact question using approximate master equations to study the coevolution of group-level features and individual-level dynamics \cite{hebert2022source,stonge2024paradoxes}. However, these works considered a fixed hypergraph structure and ignored dynamical correlations between groups. In doing so, these and all other higher-order models assume that the group structure is an exogenous, fixed, and passive architecture. The framework provided here allows modelers to relax this assumption and accommodate the full-fledged dynamic character of social systems in their description. Adaptive hypergraphs can study social dynamics that is mediated by individuals that can leave and form new groups when discontented, or recruit new members when they are satisfied.

Exposure to contagions, whether of social or biological nature, are never static and rarely pairwise. Consider the complex group-level dynamics that unfolded during the COVID-19 pandemic.
Some interventions act at the individual level (e.g., social distancing, vaccination), while others at the group level (e.g., closures of specific venues).
These interventions can create unintended consequences, as individuals may seek open groups without restrictions \cite{althouse2023unintended}. In the context of social contagions, online groups have been closed to curb the spread of hate speech, for instance banning certain subreddits on Reddit \cite{chandrasekharan2017you, trujillo2021echo}. While some users discontinue their usage of the platform, some relocate their activity to other groups. These adaptive group-level mechanisms can thus influence the spread of a contagion in complex and sometimes counterintuitive ways.

We hope that our contributions will inspire further work on the dynamics of adaptive higher-order systems.\\


{\small
\section*{Methods}

\subsection*{Equivalence between node-centered and group-centered dynamics for linear kernels}

The difference between node- and group-centered dynamics resides in the way the effective rates $\bar{\beta}_{n,i}$, $\bar{\alpha}_{n,i}$, $\tilde{\beta}_{m,l}$, and $\tilde{\alpha}_{m,l}$, are computed (regardless of whether the structure is static or adaptive). Therefore, proving that these effective rates take the same values in node- and group-centered dynamics is sufficient to demonstrate that the two are equivalent. In the following, we show this in the case in which the infection kernels are the same linear function of the number of infected nodes, namely, $\beta(k,\ell) = \delta \ell$ for node-centered dynamics and $\lambda(n,i) = \delta i$ for group-centered dynamics. Recovery rates are considered to be constant, yet an identical proof holds for recovery kernels which are the same linear function of the number of infected nodes.

In the group-centered dynamics, the effective rates $\bar{\beta}_{n,i}$ and $\tilde{\beta}_{m,l}$ take the following form,
\begin{align*}
    \bar{\beta}_{n,i} &= \delta i + \delta\left\{\avg{i}_{K_S^{i < \bar{i}}} \left[\frac{\partial}{\partial x}G_S^i(x,y)\right]_{x,y=1} + \avg{i}_{K_S^{i \geqslant \bar{i}}}\left[\frac{\partial}{\partial y}G_S^i(x,y)\right]_{x,y=1}\right\}  \ ,\\
    \tilde{\beta}_{m,l} & = \delta \left[(m-l)\avg{i}_{K_S^{i < \bar{i}}} + l \avg{i}_{K_S^{i \geqslant \bar{i}}}\right]_{x,y=1} \ .
\end{align*}

In the node-centered dynamics, we have
\begin{align*}
    \bar{\beta}_{n,i} &= \delta i + \delta \sum_{r,s} s P(r,s\vert n,i,S) = \delta i + \delta \left[\frac{\partial}{\partial y} E_S^i(x,y)\right]_{x,y=1} \\
    &= \delta i + \delta \left[\frac{\partial}{\partial y} G_S^i\left(K_S^{i < \bar{i}}(x,y),K_S^{i \geqslant \bar{i}}(x,y)\right) \right]_{x,y=1} \\
    &= \delta i + \delta\left\{\left[\frac{\partial}{\partial y}K_S^{i < \bar{i}}(x,y)\right]_{x,y=1} \left[\frac{\partial}{\partial x}G_S^i(x,y)\right]_{x,y=1} \right. \\
    &\ \ \ \ \ \ \ \ \ \ \ \ \ \ \ \ \left.+ \left[\frac{\partial}{\partial y}K_S^{i \geqslant \bar{i}}(x,y)\right]_{x,y=1} \left[\frac{\partial}{\partial y}G_S^i(x,y)\right]_{x,y=1}\right\} \ ,\\
    \tilde{\beta}_{m,l} &= \delta \sum_{k,\ell} \ell P(k,\ell\vert m,l,S) = \delta \left[\frac{\partial}{\partial y} E_S^{m,l}(x,y)\right]_{x,y=1} \\
    &= \delta \left[\frac{\partial}{\partial y} \left[\left(K_S^{i < \bar{i}}(x,y)\right)^{m-l} \left(K_S^{i \geqslant \bar{i}}(x,y)\right)^l\right] \right]_{x,y=1} \\
    &= \delta\left\{(m-l)\left[\frac{\partial}{\partial y}K_S^{i < \bar{i}}(x,y)\right]_{x,y=1} + l\left[\frac{\partial}{\partial y}K_S^{i \geqslant \bar{i}}(x,y)\right]_{x,y=1}\right\} \ ,
\end{align*}
where we used $K_S^{i < \bar{i}}(1,1) = K_S^{i \geqslant \bar{i}}(1,1) = 1$. The two equations for $\tilde{\beta}_{n,i}$ and $\tilde{\beta}_{m,l}$ coincide once observed that $\avg{i}_{K_S^i} = \left[\frac{\partial}{\partial y}K_S^i(x,y)\right]_{x,y=1}$, with either $i < \bar{i}$ or $i \geqslant \bar{i}$.

More generally, a similar proof holds also when infection and recovery kernels are linear on both variables, e.g., $\beta(k,\ell) = \xi k + \delta \ell$ and $\lambda(n,i) = \xi (n-1) + \delta i$.
}

{\small
\subsection*{Computing the effective transition rates}

For the node-centered dynamics, the effective transition rates $\bar{\alpha}_{n,i}$, $\bar{\beta}_{n,i}$, $\tilde{\alpha}_{m,l}$ and $\tilde{\beta}_{m,l}$ require that we extract the joint distributions from the bivariate PGFs $E_I^i$, $E_S^i$, $E_I^{m,l}$ and $E_S^{m,l}$, respectively.
All of them, however, are either products or compositions of functions.
By definition, for a PGF
\begin{align*}
    A(x,y) = \sum_{n = 0}^{n_\mathrm{max}} \sum_{m = 0}^{m_\mathrm{max}} a_{n,m} x^n y^m \;,
\end{align*}
we can extract the coefficient using
\begin{align*}
    a_{n,m} = \left . \frac{1}{n!m!}\frac{\partial^n\partial^m}{\partial x^n \partial y^m} A(x,y) \right |_{x,y = 0} \;.
\end{align*}
For product or composition of functions, this becomes impractical, as the number of terms for a partial derivative of degree $n$ requires $B_n$ terms, where $B_n$ is the Bell number ($B_n= 1, 2, 5, 15, 52, 203, 877, 4140, 21147, 115975\dots$ for $n = 1, 2, 3, 4, 5, 6, 7, 8, 9, 10, \dots$).

Instead, we can use the characteristic function $\phi_{n,m}(u,v)$ for the random variables $n,m$, which can be written
\begin{align*}
    \phi_{n,m}(u,v) &= \avg{e^{-2\pi i u n} e^{-2\pi i v m}} \\
                    &= \sum_{n = 0}^{n_\mathrm{max}} \sum_{m = 0}^{m_\mathrm{max}} a_{n,m} e^{-2 \pi i u n } e^{- i 2 \pi v m} \\
                    &= A(e^{-i 2 \pi u}, e^{-i 2 \pi v}) \;.
\end{align*}
The coefficients $a_{n,m}$ are then recovered exactly using an inverse discrete Fourier transform,
\begin{align*}
    a_{n,m} &= \frac{1}{(n_\mathrm{max}+1)(m_\mathrm{max}+1)}\sum_{u = 0}^{n_\mathrm{max}} \sum_{v = 0}^{m_\mathrm{max}} \phi_{n,m}(u,v) e^{2 \pi i u n} e^{i 2 \pi v m} \;.
\end{align*}
}

{\small
\subsection*{Neglecting dynamical correlations: Heterogeneous mean-field theory}

To assess the consequences of neglecting local dynamical correlations, we report here a heterogeneous mean-field (HMF) approximation and compare its predictions with GAME's. Although accounting for the membership distribution $g_m$ and the group size distribution $p_n$, the model assumes that $C_{n,i}$ is binomially distributed. Let us indicate with $I_m(t)$ ($S_m(t)$) the probability at time $t$ that a node has membership $m$ and is infected (susceptible), so that $I_m(t) + S_m(t) = g_m $. Then, depending on whether the dynamics is node- or group-centered, respectively, the HMF model is defined by the following set of equations (one for each $m$):
\begin{align*}
    \frac{dI_m(t)}{dt}\!=&\!\sum_{n_1,\dots,n_m} \sum_{i_1,\dots,i_m} \left[\prod_{k=1}^m p_{n_k}\! \sum_{i_k=1}^{n_k-1} \binom{n_k-1}{i_k}q(t)^{i_k}(1-q(t))^{n_k-1-i_k}\right] \\
    &\times\left[\beta\left(\sum_{k=1}^m n_k,\sum_{k=1}^m i_k\right)S_m(t)-\alpha\left(\sum_{k=1}^m n_k,\sum_{k=1}^m i_k\right)I_m(t)\right] \; , \\
    \frac{dI_m(t)}{dt}\!=&~m\sum_n p_n\! \sum_{i=1}^{n-1}\binom{n-1}{i}q(t)^i(1-q(t))^{n-1-i} \\
    &\ \ \ \ \ \ \ \ \ \ \ \ \ \ \ \ \ \ \ \ \ \ \ \ \ \ \ \ \ \ \ \ \ \ \times\left[\beta(n,i)S_m(t)-\alpha(n,i)I_m(t)\right]\; ,
\end{align*}
where $q(t) = \sum_{m}mI_m(t)/\sum_m m g_m = \sum_{m}mI_m(t)/\avg{m}$ is the probability of drawing an infected node.

The case where recovery is a spontaneous process (occurring at rate 1, without loss of generality) is recovered by setting $\alpha = 1/m$, to eventually get
\begin{align*}
    \frac{dI_m(t)}{dt} =&-I_m(t) + S_m(t)\sum_{n_1,\dots,n_m} \sum_{i_1,\dots,i_m} \beta\left(\sum_{k=1}^m n_k,\sum_{k=1}^m i_k\right) \\
    &\ \ \ \ \ \ \ \ \ \ \times\left[\prod_{k=1}^m p_{n_k} \sum_{i_k=1}^{n_k-1} \binom{n_k-1}{i_k}q(t)^{i_k}(1-q(t))^{n_k-1-i_k}\right] \; , \\
    \frac{dI_m(t)}{dt} =&-I_m(t) + S_m(t)~m\sum_n p_n \\
    &\ \ \ \ \ \ \ \ \ \ \ \ \ \ \ \ \ \ \ \ \times \sum_{i=1}^{n-1}\binom{n-1}{i}q(t)^i(1-q(t))^{n-1-i} \beta(n,i) \; .
\end{align*}
}

\section*{Data availability}
No datasets were generated during the current study. Empirical data used to produce the results in Fig.~\ref{fig:real_world} and Fig.~S3 of the SM are available at Ref.~\cite{netzschleuder} and Ref.~\cite{sociopatterns}, respectively.

\section*{Code availability}
The code to reproduce the results in this manuscript and its SM are available at \url{https://github.com/giubuig/GAME-Generalized-Approximate-Master-Equations}.

\begin{acknowledgments}
The authors thank Antoine Allard and Alex Arenas for discussions and feedback. We acknowledge financial support from the European Union's Horizon 2020 research and innovation program under the Marie Sk\l{}odowska-Curie Grant Agreement No.\ 945413 (G.B.), from the Universitat Rovira i Virgili (G.B.), and from the National Science Foundation award \#2242829 (G.B.); from the Fonds de recherche du Québec -- Nature et technologies \#313475 (G.S.-O.); and from the National Institutes of Health Center of Biomedical Research Excellence Award P20GM125498 (L.H.-D.).
\end{acknowledgments}

\section*{Author Contributions Statement}
G.B., G.S.-O., and L.H.-D. contributed to the development and analysis of the model, as well as to the writing and revision of the manuscript. G.B. extracted the empirical hypergraphs. G.B. and G.S.-O. performed the numerical simulations.

\section*{Competing Interests Statement}
The authors declare no competing interests.

\clearpage
\onecolumngrid

\renewcommand{\thesection}{S\arabic{section}}
\setcounter{section}{0}
\renewcommand{\theequation}{S\arabic{equation}}
\setcounter{equation}{0}
\renewcommand{\thefigure}{S\arabic{figure}}
\setcounter{figure}{0}

\renewcommand{\theHfigure}{S\arabic{figure}} 
\renewcommand{\theHequation}{S\arabic{equation}} 
\renewcommand{\theHsection}{S\arabic{section}} 

\centerline{\bf \Large Supplemental Material to}
\vspace{1em}
\centerline{\bf \Large Characteristic scales and adaptation in higher-order contagions}

\section{Generalized Approximate Master Equations (GAME)}

Recall that the GAME on static hypergraphs involves the following transitions
\begin{subequations}
\label{eq:master_equations_SM}
\begin{align}
    \frac{\mathrm{d}C_{n,i}}{\mathrm{d}t} =& \;\bar{\alpha}_{n,i+1} (i+1) C_{n,i+1} - \bar{\alpha}_{n,i} i C_{n,i}  + \bar{\beta}_{n,i-1} (n-i+1) C_{n,i-1} - \bar{\beta}_{n,i} (n-i) C_{n,i} \label{eq:master_equations_cni_SM} \; \\
    \frac{\mathrm{d} S_{m,l}}{\mathrm{d} t} =& \; \tilde{\alpha}_{m,l} I_{m,l} - \tilde{\beta}_{m,l} S_{m,l}  + \theta_S \left [ (m-l + 1) S_{m,l-1} - (m-l) S_{m,l} \right ]  + \phi_S \left [ (l + 1) S_{m,l+1} - l S_{m,l} \right ] \;, \label{eq:master_equations_sml_SM} \\
    \frac{\mathrm{d} I_{m,l}}{\mathrm{d} t} =& -\tilde{\alpha}_{m,l} I_{m,l} + \tilde{\beta}_{m,l} S_{m,l} + \theta_I \left [ (m-l + 1) I_{m,l-1} - (m-l) I_{m,l} \right ] + \phi_I \left [ (l + 1) I_{m,l+1} - l I_{m,l} \right ] \;, \label{eq:master_equations_iml_SM}
\end{align}
\end{subequations}
where the four mean fields are calculated as
\begin{subequations}
\label{eq:mf_SM}
\begin{align}
    \theta_S &= \frac{\sum_{n} (n-\bar{i}+1) (n-\bar{i}) C_{n,\bar{i}-1} \bar{\beta}_{n,\bar{i}-1}}{\sum_{n,i \leqslant \bar{i}-1} (n-i) C_{n,i}} \;, \label{eq:mf_th_s_SM}\\
    \phi_S &= \frac{\sum_{n} (n-\bar{i}) \bar{i} C_{n,\bar{i}} \bar{\alpha}_{n,\bar{i}}}{\sum_{n,i>\bar{i}-1} (n-i) C_{n,i}} \;, \label{eq:mf_ph_s_SM}\\
    \theta_I &= \frac{\sum_{n} \bar{i} (n-\bar{i}) C_{n,\bar{i}} \bar{\beta}_{n,\bar{i}}}{\sum_{n,i \leqslant \bar{i}} i C_{n,i}} \;, \label{eq:mf_th_i_SM}\\
    \phi_I &= \frac{\sum_{n} (\bar{i}+1) \bar{i} C_{n,\bar{i}+1} \bar{\alpha}_{n,\bar{i}+1}}{\sum_{n,i>\bar{i}} i C_{n,i}} \;. \label{eq:mf_ph_i_SM}
\end{align}
\end{subequations}

\section{Adaptive Generalized Approximate Master Equations (A-GAME)}

For a general adaptive hypergraphs model, we allow both susceptible and infectious nodes to rewire away from groups, respectively at rates $\gamma_S$ and $\gamma_I$. Notice that the annealed calculation of the mean fields in Eqs.~(\ref{eq:mf_SM}) remains the same, since the memory-less system does not capture dynamical correlations due to recent rewiring effects. Rewiring thus only requires the addition of the following terms (tagged with the `$\mathrm{a}$' superscript) to the static GAME,
\begin{subequations}
\label{eq:rewiring_SM}
\begin{align}
    \frac{\mathrm{d}C^\mathrm{a}_{n,i}}{\mathrm{d}t} =& \; \gamma_S {\bf 1}_{i \geqslant \bar{i}} \left[(n+1-i)C_{n+1,i} - (n-i)C_{n,i}\right] + \gamma_I \left[{\bf 1}_{i \geqslant \bar{i}} (i+1)C_{n+1,i+1} - {\bf 1}_{i > \bar{i}} iC_{n,i}\right] \notag \\
            &+ \left( \frac{\eta {\bf 1}_{i < \bar{i}}}{C_{i<\bar{i}}} + 1-\eta \right) \left[\gamma_S \Omega_{S\vert i\geqslant \bar{i}}C_{n-1,i} - \left( \gamma_S \Omega_{S\vert i\geqslant \bar{i}} + \gamma_I \Omega_{I\vert i>\bar{i}}\right)C_{n,i} \right] + \left( \frac{\eta {\bf 1}_{i \leqslant \bar{i}}}{C_{i<\bar{i}}} + 1-\eta \right) \gamma_I \Omega_{I\vert i>\bar{i}} C_{n-1,i-1} \label{eq:rewiring_cni_SM} \;,
           \\ 
    \frac{\mathrm{d} S^\mathrm{a}_{m,l}}{\mathrm{d} t} =& \; \gamma_S \left[\eta + (1-\eta) C_{i<\bar{i}}\right] \left[(l+1)S_{m,l+1}-lS_{m,l}\right] \notag \\
            &+ \gamma_I \Omega_{I\vert i>\bar{i}} \left( \frac{\eta}{\Omega_{S\vert i<\bar{i}}} + \frac{1-\eta}{\Omega_{S}} \right) \Omega_{S\vert i=\bar{i}-1} \left[(m-l+1)S_{m,l-1}-(m-l)S_{m,l}\right] \;, \label{eq:rewiring_sml_SM} 
            \\
    \frac{\mathrm{d} I^\mathrm{a}_{m,l}}{\mathrm{d} t} =& \; \gamma_I \left[\eta + (1-\eta) C_{i<\bar{i}}\right] \left[(l+1)I_{m,l+1}-lI_{m,l}\right] \notag \\
            &+ \gamma_I ~\frac{\Gamma_{I\vert i=\bar{i}+1}}{\Omega_{I\vert i>\bar{i}}} \left[(l+1)I_{m,l+1}-lI_{m,l}\right] + \gamma_I \Omega_{I\vert i>\bar{i}} (1-\eta) \frac{\Omega_{I\vert i=\bar{i}}}{\Omega_{I}} \left[(m-l+1)I_{m,l-1}-(m-l)I_{m,l}\right] \;. \label{eq:rewiring_iml_SM}
\end{align}
\end{subequations}
These terms are simply added to the previous system of equations. We defined $C_{i<\bar{i}} = \sum_{n,i<\bar{i}} C_{n,i}$, $\Omega_S = \sum_{n,i} (n-i) C_{n,i}$, $\Omega_{S \vert i \geqslant \bar{i}} = \sum_{n,i\geqslant \bar{i}} (n-i) C_{n,i}$, $\Omega_{S \vert i < \bar{i}} = \sum_{n,i < \bar{i}} (n-i) C_{n,i}$, $\Omega_{S\vert i = \bar{i}-1} = \sum_{n} (n-\bar{i}+1) C_{n,\bar{i}-1}$, $\Omega_I = \sum_{n,i} i C_{n,i}$, $\Omega_{I\vert i > \bar{i}} = \sum_{n,i > \bar{i}} i C_{n,i}$, $\Omega_{I\vert i = \bar{i}} = \sum_{n} \bar{i} C_{n,\bar{i}}$, and $\Gamma_{I\vert i = \bar{i}+1} = \sum_{n} \bar{i}(\bar{i}+1) C_{n,\bar{i}+1}$. The first row in Eq.~(\ref{eq:rewiring_cni_SM}) account for nodes leaving groups, the second for nodes joining groups. The first term in Eq.~(\ref{eq:rewiring_sml_SM}) accounts for S nodes rewiring away, while the second term for I nodes rewiring to groups with $i=\bar{i}-1$, making them active for the S nodes therein. Lastly, the first term in Eq.~(\ref{eq:rewiring_iml_SM}) accounts for I nodes rewiring away, while the second (third) term accounts for I nodes leaving (joining) groups with $i=\bar{i}+1$ ($i=\bar{i}$), making them inactive (active) for the I nodes therein. The simpler version where only S nodes rewire is recovered by taking $\gamma_I=0$.

\section{GAME: Additional results}

We provide here supplemental results to the main text (i) on the assessment of the characteristic scale $\bar{i}$ that best captures the contagion process simulated via Monte Carlo runs; and (ii) to test the accuracy of the GAME for node-based dynamics while also comparing it with the predictions from node-centered AME~\cite{marceau2010adaptive}. 

\begin{figure}
    \centering
    \includegraphics[width=\linewidth]{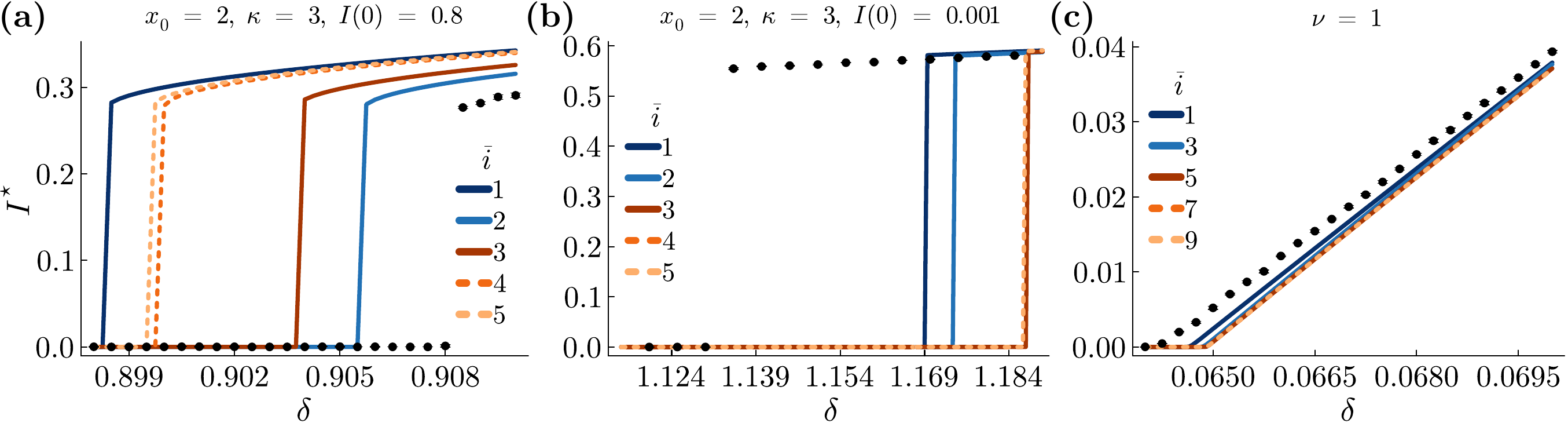}
    \caption{Comparison of the results for the equilibrium prevalence, $I^\star$, obtained on random 3-regular hypergraphs integrating the model and via Monte Carlo simulations, for the values of $\bar{i}$ indicated in the legends. (a \& b) Results for a sigmoidal infection kernel of the form $\lambda(n,i) = \delta~[s(i; x_0, \kappa) - s(0; x_0, \kappa)]/s(n_\text{max}; x_0, \kappa)$, where $s(i; x_0, \kappa) = 1/[1+e^{-\kappa (i - x_0)}]$. The group size distribution follows a truncated Poisson with mode $n=4$ and support $\{2,\dots,n_\text{max} = 6\}$. Results around (a) the persistence threshold ($I(0) = 0.8$) and (b) the invasion threshold ($I(0) = 0.001$). (c) Results for a linear infection kernel in a regime of mesoscopic localization, obtained considering a bimodal group size distribution where $99\%$ of groups have size $n=4$ and the remainder $n=15$.
    Solid and dashed lines represent the predictions provided by the model, while points and error bars (when visible) denote averages and standard errors over the last $200$ time-steps of a Monte Carlo simulation performed on hypergraphs of (a \& b) $N=10^6$ and (c) $N=5\times 10^6$ nodes.
    }
    \label{fig:dyn_corr}
\end{figure}

Figures~\ref{fig:dyn_corr}(a) and (b) consider a steep sigmoidal infection kernel with inflection point at $i = x_0$, showing how $\bar{i} = x_0$ is the most appropriate scale ($x_0 = 2$ in the reported example). The only exception to this regards the invasion threshold, which is better located by setting $\bar{i}=1$, as we see in Fig.~\ref{fig:dyn_corr}(b). This can be expected, for most infectious groups in a nearly susceptible population contain just one infected node. Nonetheless, once above the threshold predicted by $\bar{i} = x_0$, the latter is the most accurate model.

Figure~\ref{fig:dyn_corr}(c) considers instead a linear infection kernel but in a regime of mesoscopic localization, showing how, as in the delocalized regime considered in Fig.~5 of the main text, the characteristic scale coincides with $\bar{i} = 1$ (or with $\bar{i} = \nu$, for a more general threshold-like kernel proportional to ${\bf 1}_{i \geqslant \nu} i$).

\begin{figure*}
    \centering
    \includegraphics[width=\linewidth]{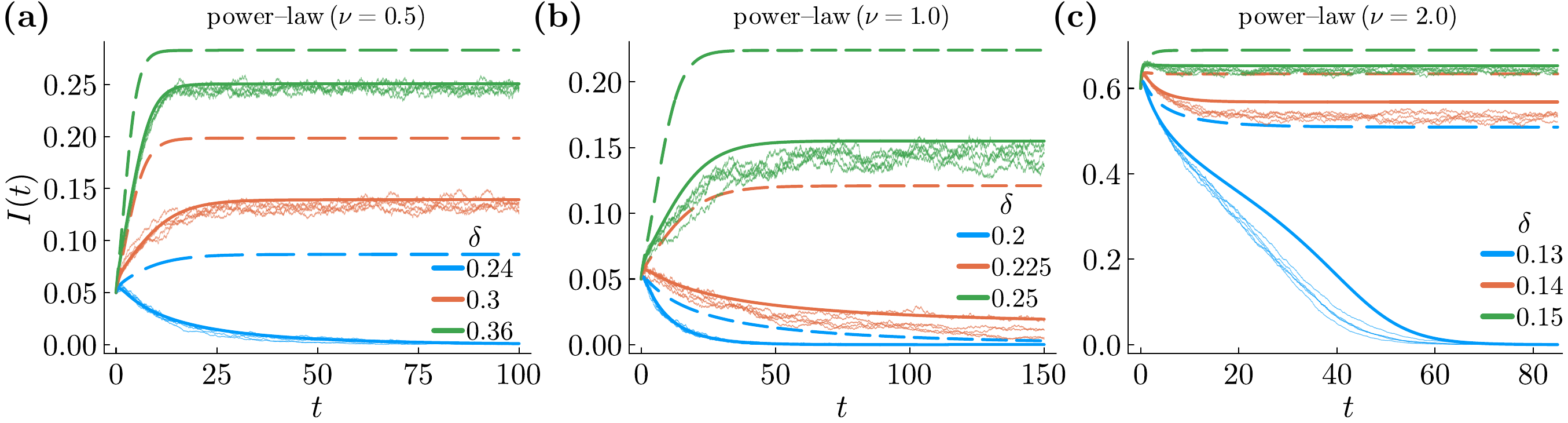}
    \caption{Comparison between GAME's (solid) and node-based AME's (dashed) predictions against Monte Carlo simulations under node-based dynamics. We use 2-regular 4-uniform hypergraphs of $N = 3\times 10^4$ nodes. Five trajectories from Monte Carlo simulations are reported for each value of the infection rate $\delta$. Power-law kernels $\beta(n,i) = \delta i^\nu$ for (a) $\nu = 0.5$ ($I(0) = 0.05$), (b) $\nu = 1.0$ ($I(0) = 0.05$), and (c) $\nu = 2.0$ ($I(0) = 0.6$). The GAME uses $\bar{i} = 1$.}
    \label{fig:node}
\end{figure*}

Lastly, in Fig.~\ref{fig:node} we compare some temporal evolutions from Monte Carlo simulations to those predicted by either the GAME or the node-based AME under node-centered dynamics. As for group-centered dynamics, where the GAME (at optimal $\bar{i}$) substantially improves on group-based AME (i.e., the GAME at $\bar{i} \geq n_\text{max}$; see Fig.~5 of the main text and Fig.~\ref{fig:dyn_corr}), the GAME also proves to be highly accurate under node-centered dynamics, significantly improving on node-based AME.

\section{Heterogeneous mean-field theory and comparison with the GAME}

We report here the HMF equations presented in the Methods section of the main text and additional results to compare the HMF to the GAME. The dynamical equations read
\begin{align}
    \frac{dI_m(t)}{dt} =&~m\sum_n p_n \sum_{i=1}^{n-1}\binom{n-1}{i}q(t)^i(1-q(t))^{n-1-i} \left[\beta(n,i)S_m(t)-\alpha(n,i)I_m(t)\right]\ \ \ \ \ \ \textrm{(group-centered)}; \label{eq:HMF_group_gen}\\
    \notag \frac{dI_m(t)}{dt} =& \sum_{n_1,\dots,n_m} \sum_{i_1,\dots,i_m} \left[\prod_{k=1}^m p_{n_k} \sum_{i_k=1}^{n_k-1} \binom{n_k-1}{i_k}q(t)^{i_k}(1-q(t))^{n_k-1-i_k}\right] \times \\
&\ \ \ \ \ \ \ \ \ \ \ \ \ \ \ \ \ \ \ \ \ \ \ ~\times\left[\beta\left(\sum_{k=1}^m n_k,\sum_{k=1}^m i_k\right)S_m(t)-\alpha\left(\sum_{k=1}^m n_k,\sum_{k=1}^m i_k\right)I_m(t)\right] \ \ \ \ \ \ \textrm{(node-centered)}; \label{eq:HMF_node_gen}
\end{align}
where $q(t) = \sum_{m}mI_m(t)/\sum_m m g_m = \sum_{m}mI_m(t)/\avg{m}$ is the probability of drawing an infected node.

The case where recovery is a spontaneous process (occurring at rate 1, without loss of generality) is recovered by setting $\alpha = 1/m$, to eventually get
\begin{align}
    \frac{dI_m(t)}{dt} =&-I_m(t) + S_m(t)~m\sum_n p_n \sum_{i=1}^{n-1}\binom{n-1}{i}q(t)^i(1-q(t))^{n-1-i} \beta(n,i)\ \ \ \ \ \ \textrm{(group-centered)}; \label{eq:HMF_group}\\
    \frac{dI_m(t)}{dt} =&-I_m(t) + S_m(t)\sum_{n_1,\dots,n_m} \sum_{i_1,\dots,i_m} \left[\prod_{k=1}^m p_{n_k} \sum_{i_k=1}^{n_k-1} \binom{n_k-1}{i_k}q(t)^{i_k}(1-q(t))^{n_k-1-i_k}\right] \beta\left(\sum_{k=1}^m n_k,\sum_{k=1}^m i_k\right) \ \ \ \ \ \ \textrm{(node-centered)}. \label{eq:HMF_node}
\end{align}

\begin{figure}
    \centering
    \includegraphics[width=\linewidth]{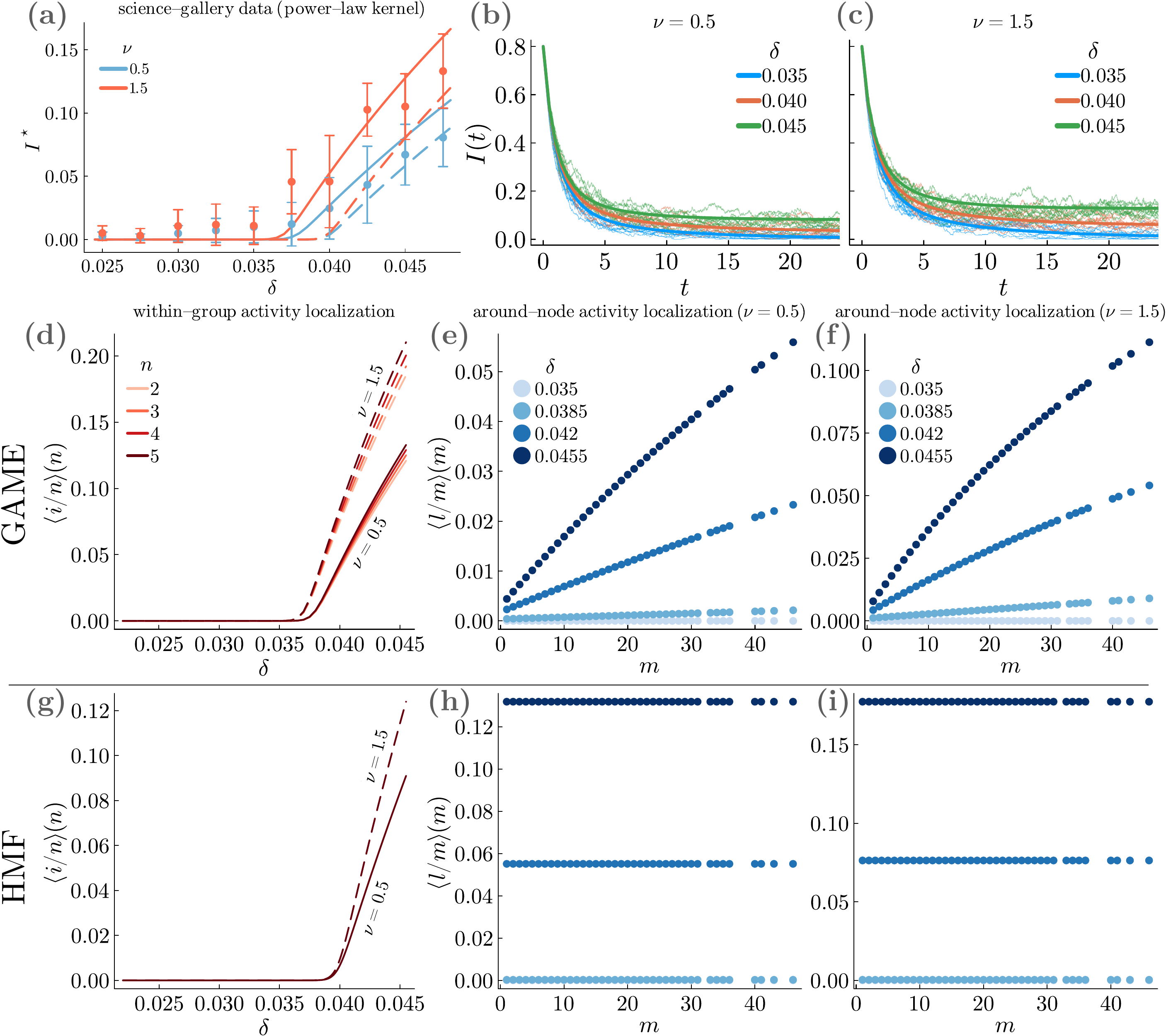}
    \caption{Results for a group-centered dynamics on a hypergraph built from face-to-face interactions data~\cite{sociopatterns}. A power-law kernel $\beta(n,i) = \delta i^\nu$ with $\nu=0.5,1.5$ is used. (a) Comparison of the GAME's (solid) and HMF theory's (dashed) predictions with Monte Carlo simulations. Points and error bars respectively represent averages and standard errors over 20 random realizations. (b-c) Temporal evolution for the fraction $I(t)$ of infected nodes as predicted by the GAME (solid) against 5 trajectories from Monte Carlo simulations for each value of $\delta$ indicated in the legend. (d-i) Activity localization as predicted by (d-f) the GAME and (g-i) the HMF theory (Eq.~(\ref{eq:HMF_group})). In (g) the four curves are superposed; the theory is unable to capture size-dependent activity.}
    \label{fig:HMF&GAME_sciencegallery}
\end{figure}

An immediate consequence of getting rid of the group state distribution ($C_{n,i}$) is the inability to predict the dependence of the invasion threshold (if present) on higher-order infection channels~\cite{burgio2024triadic,burgio2021network}. In fact, around the contagion-free state, i.e., $I\in \mathcal{O}(\epsilon)$ with $\epsilon \ll 1$, only the terms with $i=1$ survive at order $\epsilon$, so that all contributions to infection mediated by two or more nodes appear to be irrelevant in the HMF approach, even though they are not.
For example, the HMF theory incorrectly predicts the same threshold for power-law kernels with different exponents $\nu$. Overall, it generally provides significantly less accurate predictions than the GAME; see for instance Fig.~4 of the main text. As expected, predictions become comparable when very dense structures are considered. This can be observed in Fig.~\ref{fig:HMF&GAME_sciencegallery}(a), where we used a dense hypergraph built from face-to-face interaction data recorded in a science gallery~\cite{sociopatterns} (see Sec.~\ref{sec:real_world} for details). 



One key reason for which previous models and especially HMF theories can be substantially less accurate than the GAME is that, by neglecting some or all local dynamical correlations, they are unable to properly capture the variation in activity between groups of different sizes and/or between nodes of different memberships. This is exemplified in Fig.~\ref{fig:HMF&GAME_sciencegallery}(d)-(i), where we report the expected fraction $\avg{i/n}(n)$ of active nodes in groups of size $n$ and the expected fraction $\avg{l/m}(m)$ of active groups incident on a node of membership $m$, for the science-gallery data hypergraph. The GAME capture the activity localization either within larger groups---$\avg{i/n}(n)$ increases with $n$---and around more connected nodes---$\avg{l/m}(m)$ increases with $m$. In contrast, HMF theory incorrectly predicts that $\avg{i/n}(n)$ and $\avg{l/m}(m)$ are independent of $n$ and $m$, respectively. Group-based AME~\cite{hebert2010propagation,st-onge2021master}, on the other hand, can reproduce the localization within groups but not the localization around nodes ($\avg{l/m}(m)$ is predicted to be independent of $m$), as they do not track the activity around the latter.

\begin{figure}
    \centering
    \includegraphics[width=\linewidth]{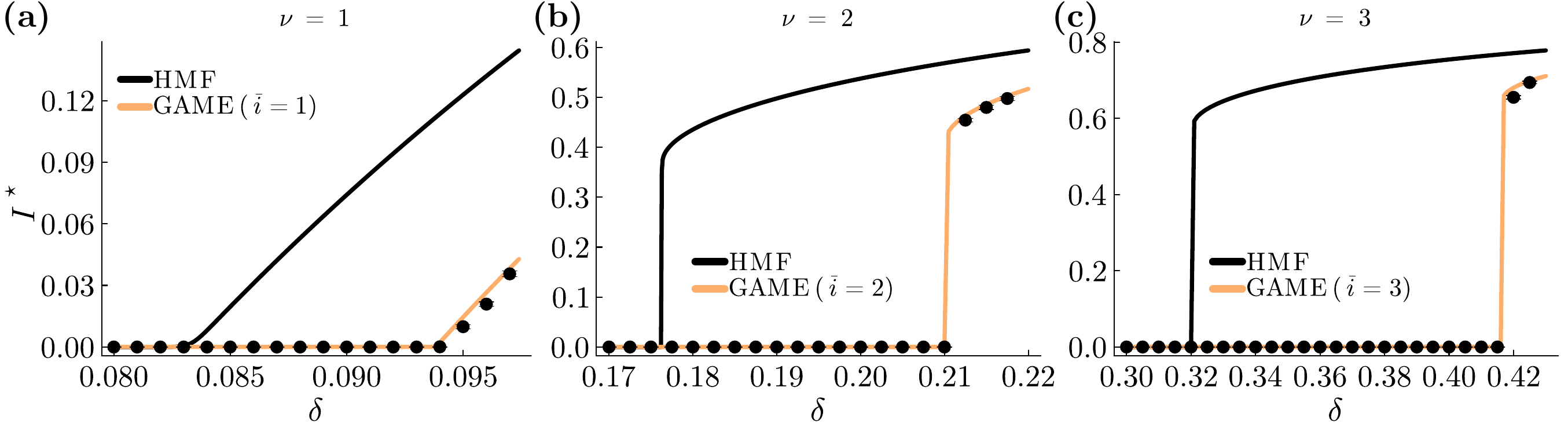}
    \caption{Equilibrium prevalence, $I^\star$, obtained on random 3-regular 5-uniform hypergraphs under group-centered dynamics considering a threshold infection kernel $\lambda(n,i) = {\bf 1}_{i \geqslant \nu} \delta i$, for (a) $\nu=1$ ($I(0)=0.06$), (b) $\nu=2$ ($I(0)=0.8$) and (c) $\nu=3$ ($I(0)=0.8$). Black and orange solid lines represent the results obtained integrating the HMF model and the (best) GAME model, respectively; points and error bars (when visible) denote averages and standard errors over $20$ random realizations resulting from Monte Carlo simulations performed on hypergraphs with $N=5\times 10^4$ nodes. The HMF predicts the critical threshold with a relative error of around $10\%$ for $\nu = 1$, $16\%$ for $\nu = 2$, and $23\%$ for $\nu = 3$. The GAME reduces those errors to values not larger than, respectively, $0.02\%$, $0.1\%$, and $0.1\%$; that is, approximately by a factor of $100$ to $500$.}
    \label{fig:HMF&GAME_threshold}
\end{figure}

Even when activity localization cannot occur, as---by construction---in regular and uniform hypergraphs, keeping local dynamical correlations is anyway crucial. As shown in Fig.~\ref{fig:HMF&GAME_threshold}, the accuracy of the HMF theory can indeed drop drastically just because of structural sparsity.

\section{A-GAME: Additional results}

\begin{figure}[t]
    \centering
    \includegraphics[width=0.6\linewidth]{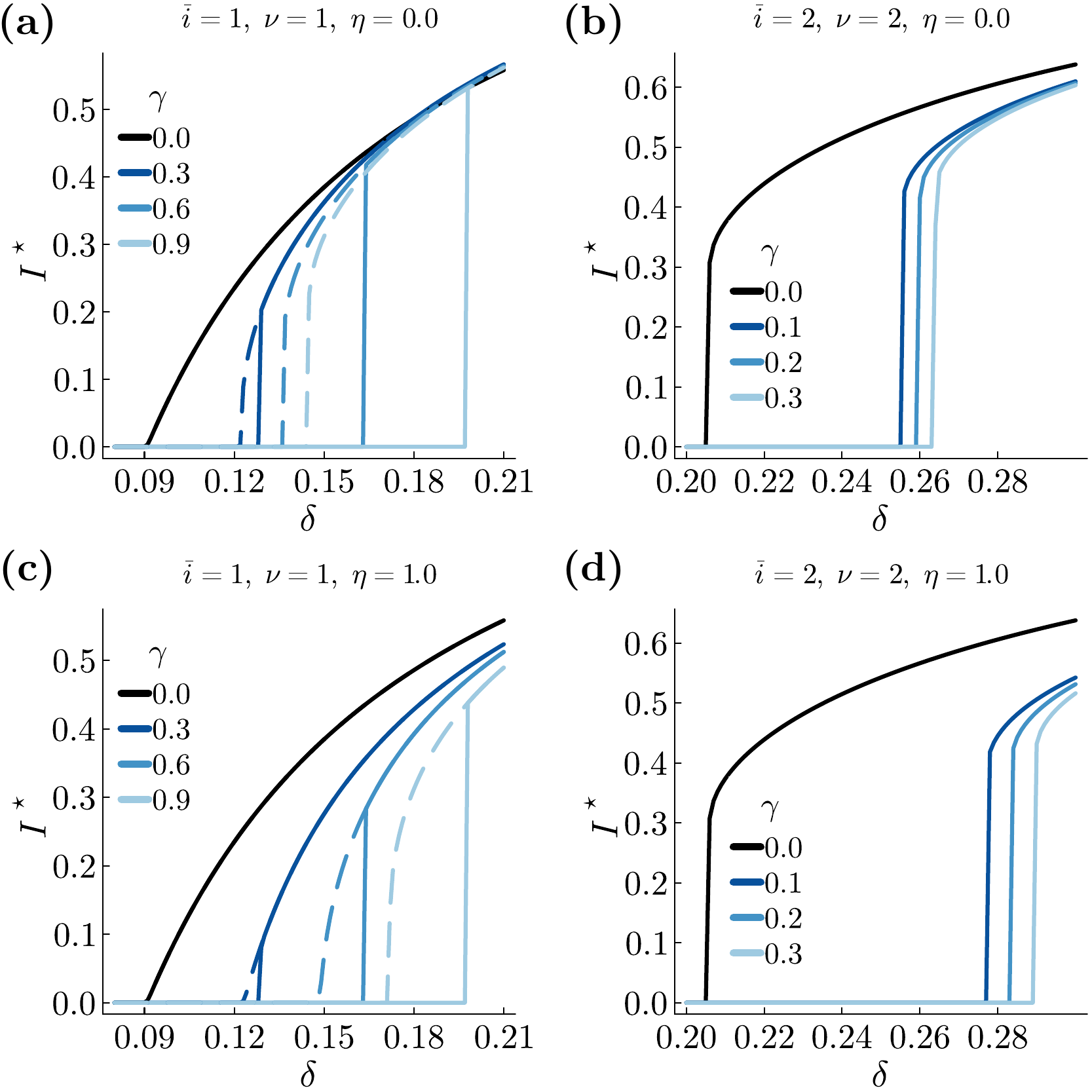}
    \caption{Equilibrium prevalence, $I^\star$, versus $\delta$ for different values of the rewiring rate, $\gamma$. As in Fig.~6 of the main text, we consider a threshold-like infection kernel $\lambda(n,i) = {\bf 1}_{i \geqslant \nu} \delta i$ to model either (a \& c) a simple contagion ($\nu = 1$) or (b \& d) a complex contagion ($\nu = 2$), while assuming the rewiring strategy $\bar{i} = \nu$. Increasing $\gamma$ widens the bistability region for $\nu = 1$ and increases the persistence threshold for $\nu = 2$. Comparing (a) to (c) and (b) to (d), respectively, we can appreciate the effect of increasing the rewiring accuracy from $\eta = 0.0$ to $\eta = 1.0$, leading the persistence threshold to increase (and the bistability region to shrink, if any).
    }
    \label{fig:adaptive_SM}
\end{figure}

We report here additional results to the main text (see Fig.~6). Considering a threshold-like infection kernel $\lambda(n,i) = {\bf 1}_{i \geqslant \nu} \delta i$, Figure~\ref{fig:adaptive_SM} shows the effects the rewiring rate, $\gamma$, and rewiring accuracy, $\eta$, have on the phase diagram. Specifically, the bistable region is widened by either increasing $\gamma$---the invasion threshold increases faster than the persistence one---or decreasing $\eta$---the invasion threshold is unaffected while the persistence one decreases. For complex contagion ($\nu > 1$), the system shows bistability as already does in the static case---rewiring moves the persistence threshold but does not produce new equilibria. Please, notice that these results are not specific to the used kernel.

\begin{figure}
    \centering
    \includegraphics[width=\linewidth]{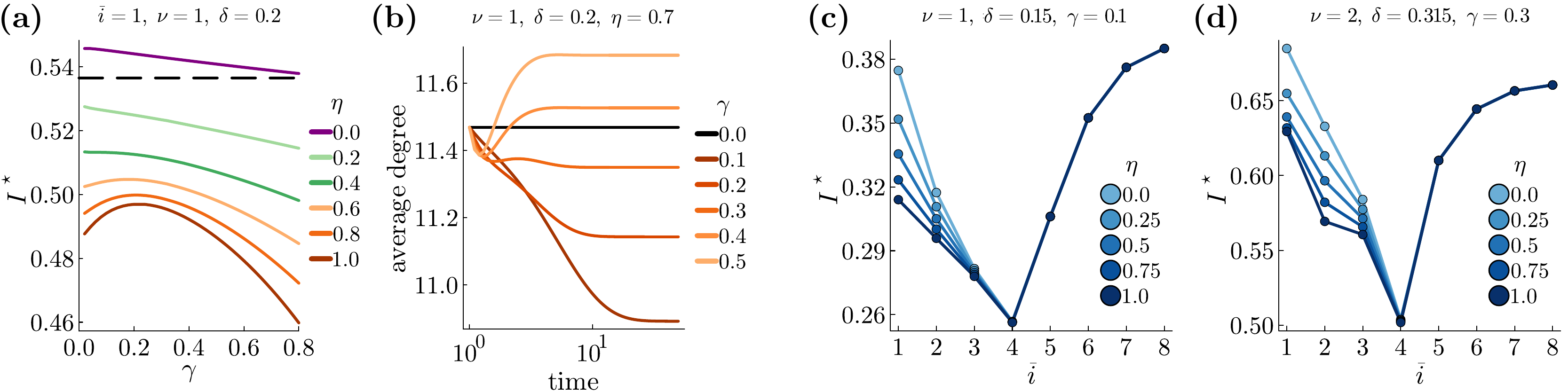}
    \caption{Further results for the adaptive hypergraphs in Fig.~6 of the main text.
    (a) Horizontal slices of the phase diagram in Fig.~6(a). The purple curve corresponds to detrimental rewiring, where the equilibrium prevalence, $I^\star$, is larger than in absence of rewiring (dashed line); green curves represent increasingly beneficial rewiring; orange curves denote non-monotonic beneficial rewiring, where $I^\star$ first increases and then decreases.
    (b) Time evolution of the average degree for different rewiring rates $\gamma$ in the non-monotonic regime, $\eta = 0.7$.
    (c \& d) For slow rewiring, it is always optimal for agents to target the structure and minimize degree by setting $\bar{i} = 4 \approx \avg{n}$, no matter the rewiring accuracy.
    }
    \label{fig:adaptive2_SM}
\end{figure}

Figure~\ref{fig:adaptive2_SM}(a) helps distinguish the different regimes determined by the rewiring accuracy, $\eta$. In particular, we can appreciate the regime of high enough accuracy where an intermediate, least-optimal rewiring rate exists (orange curves). This emerges as an intermediary case where the two best strategies, i.e., targeting the dynamics (as for high enough $\gamma$) or the structure, are implemented in the least optimal way. Note that this is not the worst possible regime, as the local maximum still outperforms dynamics without any rewiring.

The fact that, in a region of low $\gamma$, the equilibrium fraction of infected nodes, $I^\star$, decreases by lowering $\gamma$ comes from the fact that a smaller $\gamma$ ensures a lower connectivity. Indeed, for very slow rewiring, at the typical rewiring time ($\gamma^{-1}$) many nodes are already infected (in fact, the slower the rewiring, the more the fraction of infected nodes overshoots initially) and the probability that a group includes at least $\nu$ infected nodes (and thus is infectious and avoided by susceptible nodes adopting $\bar{i}=\nu$) correlates already strongly with the size of the group. Consequently, especially for high rewiring accuracy, susceptible nodes will often escape large groups and target small groups, eventually leading to a more homogeneous group size distribution, hence to a lower average degree (recall that the latter is proportional to the variance of the group size distribution). When the rewiring rate is slightly increased, that correlation becomes weaker, in turn implying a less homogeneous group size distribution, thus a larger connectivity (see Fig.~\ref{fig:adaptive2_SM}(b)). However, rewiring is still too slow for the strategy $\bar{i}=\nu$ to work. Increasing the rewiring rate further, given the accuracy is high enough, the dynamics-targeting strategy $\bar{i}=\nu$ performs better and better, up to the point at which $I^\star$ decreases again with $\gamma$. In other words, the least optimal rewiring rate is too slow to readily avoid infections but too fast to minimize the degree.

Lastly, we show in Figs.~\ref{fig:adaptive2_SM}(c) and (d) that minimizing connectivity, as implied by setting $\bar{i} \approx \avg{n}$, is always the best strategy when rewiring is slow, no matter how accurate the rewiring is.

\section{Hypergraphs from real-world interactions data}\label{sec:real_world}

\paragraph*{Board directors dataset.} This is the dataset used in Fig.~4 of the main text. It consists of records of board directors (nodes) co-sitting on common boards (hyperedges) of Norwegian public limited companies~\cite{seierstad2011few,netzschleuder}. A network is provided for each month from May 2002 to August 2011. We used the one from May 2008 (\texttt{net\char`1m\char`_2008-05-01/edges.csv}), as this is the one with the largest of the largest connected components, consisting of 870 nodes. Promoting the maximal cliques to hyperedges we end up with 219 group interactions of sizes from 3 to 13, mean group size $\avg{n} \approx 5.36$ (st. dev. $\approx 1.81$), and mean membership $\avg{m} \approx 1.35$ (st. dev. $\approx 0.88$).

\paragraph*{Science gallery dataset.} This is the dataset used in Fig.~\ref{fig:HMF&GAME_sciencegallery}. It comprehends time-resolved, face-to-face pairwise interactions collected on a daily basis during a science gallery exhibition from April 28th to July 17th, 2009~\cite{sociopatterns}. We used, in particular, the data from one of the busiest days, July 15th (\texttt{listcontacts\char`_2009\char`_07\char`_15.txt}), with 17298 time-stamped pairwise interactions involving 410 individuals. Group interactions are associated to the cliques of the observed temporal network. For instance, if three interactions involving three individuals (say, agent1-agent2, agent2-agent3, and agent1-agent3) have all the same time stamp, they form a clique at that time. The three individuals are therefore all simultaneously interacting as a group of three. The resulting hypergraph contains 14275 hyperedges with sizes between 2 and 5. We filter it by getting rid of hyperedges which are either repeated or fully contained in others, being these not considered in the model (even though this can be easily extended by weighting interactions differently). The resulting hypergraph has 410 nodes and 2145 hyperedges of size between 2 and 5; mean group size $\avg{n} \approx 2.38$ (st. dev. $\approx 0.54$) with around 35\% of the groups having size larger than 2, and mean membership $\avg{m} \approx 12.43$ (st. dev. $\approx 8.51$). Notice that the filtering procedure does not significantly change the shape of the group size and membership distributions.

\end{document}